\documentclass[floatfix,aip,jcp,amsmath,amssymb,reprint]{revtex4-2}

\usepackage{epsfig}
\usepackage{latexsym}
\usepackage{amsmath}
\usepackage{amsfonts}
\usepackage{amssymb}
\usepackage{amsbsy}
\usepackage{subfigure}
\usepackage{bm}
\usepackage{color}
\usepackage{dcolumn}
\usepackage{hyperref}
\usepackage{graphicx}
\usepackage[utf8]{inputenc}
\usepackage[T1]{fontenc}
\usepackage{mathptmx}
\usepackage{etoolbox}

\newcommand{\lla}{\left\langle}
\newcommand{\rra}{\right\rangle}

\newcommand{\REV}[1]{\textcolor{black}{#1}}

\makeatletter
\def\@email#1#2{%
 \endgroup
 \patchcmd{\titleblock@produce}
  {\frontmatter@RRAPformat}
  {\frontmatter@RRAPformat{\produce@RRAP{*#1\href{mailto:#2}{#2}}}\frontmatter@RRAPformat}
  {}{}
}%
\makeatother
\begin{document}

\title[Active semiflexible polymer under shear flow]{Two-dimensional active polar
  semiflexible polymer under shear flow}
\author{A. Lamura}
\email{antonio.lamura@cnr.it}
\affiliation{Istituto Applicazioni Calcolo, CNR, Via Amendola 122/D,
  70126 Bari, Italy}
\author{R. G. Winkler}%
\affiliation{Theoretical Soft
Matter and Biophysics, Institute for Advanced Simulation,
Forschungszentrum J\"{u}lich, 52428 J\"{u}lich, Germany}%

\date{\today}

\begin{abstract}
The nonequilibrium structural and dynamical properties of semiflexible active polar polymers 
subject to linear flow are studied by numerical simulations. Filaments are confined in two dimensions
and immersed in a fluid described by the Brownian Multiparticle Collision Dynamics approach.
The applied shear flow causes conformational changes of a polymer, aligns it along the flow direction,
and induces a tumbling motion at large flow rates. 
In an intermediate, activity-dependent shear-rate regime, a characteristic scaling exponent for the
mean-square end-to-end distance along the gradient direction is observed. This exponent appears to be
determined by the semiflexibility of the polymer. The tumbling dynamics exhibits a characteristic time,
with a stronger dependence on the Weissenberg number than that of flexible active or passive polymers. 
\REV{Activity strongly impacts the rheological properties of the semiflexible polymers, and even implies a negative viscosity for weak flows.}
At very large values of the shear rate, shear dominates over activity and passive-polymer behavior is assumed.
\end{abstract}

\maketitle

\section{Introduction}

Autonomously moving or driven filamentous and slender-body-type structures are an integral part of biological systems. The resulting nonequilibrium, activity-induced fluctuations give raise to structural and dynamical properties, which are absent in equilibrium systems~\cite{wink:20}. On subcellular length scales, replication of DNA or transcription of RNA generates nonthermal fluctuations~\cite{albe:22,guth:99,meji:15,dipi:18}. Active processes are proposed to play an essential role in the arrangement of the eukaryotic genome and in the control of their
dynamical properties~\cite{dipi:18,jave:13,zido:13,lieb:09,gana:14,sain:18.1,wink:20,goyc:23}, \REV{as well as in the phase separation of active polymer systems~\cite{smre:17}.} 
Molecular motors walking along microtubules or actin filaments give rise to nonequilibrium conformational fluctuations~\cite{nedl:97,bran:08,webe:15}, and generate forces that affect the cytoskeletal network and organization of the cell interior~\cite{lau:03,mack:08,lu:16,ravi:17}. On a more macroscopic scale, worms such as the California blackworm~\cite{nguy:21}, {\em Caenorhabditis elegans}~\cite{gagn:16,ding:19}, or {\em T. tubifex}~\cite{debl:20,debl:20.1}, can be considered as active semiflexible filaments, which exhibit polymer characteristics and collectively organize into dynamical networks~\cite{sugi:19}. 

The presence of external flows can substantially affect the conformational, dynamical, and transport properties of active filaments~\cite{sain:18.1,pand:25,pand:25.2} as well as their rheological behavior. Experimental studies on the above worms unveil an enhanced shear-thinning behavior~\cite{debl:20,malv:19}, differing qualitatively from that of passive polymeric fluids~\cite{lars:99,xu:22}.  Theoretical approaches~\cite{wink:20,mart:18.1,wink:24,pand:25.2} and computer simulations~\cite{pand:23,pand:25,kuma:24} of various active-polymer models reveal a tight coupling between the active forces and shear flow. Thereby, different implementations of the active forces---active Brownian polymers (ABPOs)\cite{kais:14,eise:16,wink:20,anand:18}, active polar polymers (APPs)~\cite{isele:15,bian:18,pete:20,phil:22.1,phil:22,faze:23,kuma:24,teje:24,wink:24,pand:25,wink:25,kara:24}---result in comparable qualitative features as observed for passive polymers. However, the flexible active polymer models, ABPOs and APPs, exhibit major quantitative conformational, dynamical, and rheological differences. In particular, the polar active polymers exhibit an enhanced shear-thinning behavior over an activity-dependent shear-rate regime, before shear dominates over activity and the passive polymer behavior is assumed for large shear rates~\cite{pand:25,pand:25.2}.

Actin filaments and microtubules are rather semiflexible or stiff than flexible~\cite{nedl:97,sanc:12,doos:18}. The active dynamics of such  molecules is often studied via motility assays~\cite{scha:10,kawa:08,liu:11,keya:20}. To shed light onto the influence of stiffness on the properties of polar active filaments under shear flow, we perform simulations of semiflexible polymers in two dimensions. By using a coarse-grained, generic approach that mimics the self-propulsion of linear filamentous~\cite{isele:15,bian:18,anand:18,wink:20,phil:22.1,faze:23,teje:24,pand:25.2} \REV{or ring-like structures~\cite{loca:21,lamu:24_2}}, simulations reveal a significant impact of semiflexibility on the filament’s conformational and dynamical properties~\cite{isele:15,kara:24,janz:24} compared to  passive polymers~\cite{huan:10,wink:10} as well as flexible polar polymers~\cite{pand:25,pand:25.2}. Specifically, the shear-induced stretching along the flow direction is affected, as semiflexible polymers can assume compact, even spiral- or ring-like, conformations in two dimensions on account of excluded-volume interactions~\cite{isele:15,kawa:08,liu:11,keya:20}. Moderate shear resolves such structures and an activity-enhanced alignment and stretching along the flow direction occurs. At higher flow strengths, a polymer exhibits tumbling motion, where it folds into U- or S-shaped structures, which reduces its overall extension along the flow direction. This stretching is tightly linked with a shrinkage in the gradient direction due to the fixed polymer contour length. The mean-square end-to-end distance or the radius of gyration along the gradient direction of passive polymers decrease according to the power law $\dot \gamma^{-\nu}$ with the exponent $\nu \lesssim  1/2$, as the shear rate $\dot \gamma$ increases, in presence of excluded-volume interactions~\cite{schr:05.1,huan:10,lamu:12}, and  $\nu \approx 2/3$ for phantom polymers~\cite{wink:10} independent of stiffness. Our simulations of passive polymers are in agreement with these results. However, for large activities, we find a quantitatively different behavior over an activity-depended shear-rate regime, with the exponent $\nu \approx 1$. This enhanced drop with increasing shear rate is attributed to the semiflexible nature of the polymers and their conformational properties along the flow direction. The particular features of semiflexible polymers are also reflected in the tumbling motion, a cyclic stretching and recoiling dynamics~\cite{smit:99,teix:05,wink:06}. Passive polymers exhibit a characteristic time for this dynamics, denoted as tumbling time, $\tau_\phi$, which decreases approximately as $\tau_\phi \sim \dot \gamma^{-2/3}$ for large shear rates~\cite{schr:05,puli:05,wink:06,huan:10,wink:10,pinc:23}. Our simulations yield the approximate power-law $\dot \gamma^{-3/4}$ at large P\'eclet numbers in an intermediate shear-rate regime. This dependence not only differs from the passive polymer behavior, but also from that of flexible active polar polymers, where simulations and theory suggest the power-law $\dot \gamma^{-1/3}$.~\cite{pand:25,pand:25.2} Again, we attribute this enhanced dynamics to the particular conformations of the semiflexible polymer. \REV{Our simulations yield a negative contribution to viscosity for semiflexible polymers and weak shear flows. Here, the active contribution to the shear stress dominates over that by the intramolecular forces. The presence of a negative viscosity has been predicted for puller-type microswimmers due to the interference of their intrinsic flow field with the external one~\cite{hatw:04,sain:10,marc:13}. This differs from our observations, as we consider dry systems. } 
In the limit of very large shear rates, flow dominates over activity and the polymers behave as passive ones. This is consistent with the properties of flexible polymers in three dimensions~\cite{pand:25,pand:25.2}.

\section{Model and Method} \label{sec:model}

The two-dimensional wormlike filament of length $L=(N-1)r_0$ is composed of $N$ beads of mass $M$ at the positions $\bm r_i$ ($i=1,\ldots,N$) separated by the equilibrium bond length $r_0$. The \REV{intramolecular} interactions between the beads are captured by the potential   
$U=U_{bond}+U_{bend}+U_{ex}$, which includes bond, bending, and excluded-volume contributions \cite{isele:15,lamu:19,lamu:21}. Bond forces between subsequent beads are modeled by the harmonic potential
\begin{equation}\label{bond}
U_{bond}=\frac{\kappa_h}{2} \sum_{i=1}^{N-1}
(|{\bm r}_{i+1}-{\bm r}_{i}|-r_0)^2 ,
\end{equation}
with the bond strength $\kappa_h$. Restrictions in bending between \REV{subsequent} bonds are included via  the potential
\begin{equation}
U_{bend}=\kappa \sum_{i=1}^{N-2} (1-\cos \varphi_{i}) ,
\label{bend}
\end{equation}
where $\kappa$ is the bending rigidity and $\varphi_{i}$ is the angle between two adjoining  
bond vectors. Finally, the truncated and shifted Lennard-Jones potential  
\begin{equation}
U_{ex} =
4 \epsilon \Big [ \Big(\frac{\sigma}{r}\Big)^{12}
-\Big(\frac{\sigma}{r}\Big)^{6} +\frac{1}{4}\Big] \Theta(2^{1/6}\sigma -r) .
\label{rep_pot}
\end{equation}
describes the excluded-volume interactions~\cite{alle:87}.
Here, $r$ is the distance between two unbound beads,
$\epsilon$ is the interaction energy, and
$\Theta(x)$ is the Heaviside function ($\Theta(x)=0$ for $x<0$ and
$\Theta(x)=1$ for $x \ge 0$). 

The active forces are applied along the bond vectors, with the total active force on the $i^{th}$ bead  
 \cite{jiang:14,anand:18,pand:25}
\begin{equation}
{\bm F}_i^a =
\frac{f_a}{2} \left [ \frac{{\bm r}_{i+1}-{\bm r}_{i}}{|{\bm r}_{i+1}-{\bm r}_{i}|}
+   \frac{{\bm r}_{i}-{\bm r}_{i-1}}{|{\bm r}_{i}-{\bm r}_{i-1}|}      \right ] .
\label{force}
\end{equation}
The active force on the first and last bead is ${\bm F}_1^a = ({\bm r}_{2}-{\bm r}_{1})/(2 |{\bm r}_{2}-{\bm r}_{1}|)$ and ${\bm F}_N^a = ({\bm r}_{N}-{\bm r}_{N-1})/(2 |{\bm r}_{N}-{\bm r}_{N-1}|)$, respectively.
Other realizations of  the active force have \REV{also} been adopted \cite{isele:15,bian:18,phil:22_1,faze:23}. The actual choice may affect the conformational properties of the polar polymer \cite{wink:25,janz:25}. The present choice is very similar to the active force ${\bm F}_i^a =
f_a ({\bm r}_{i+1}-{\bm r}_{i-1}) / (2 r_0)$ applied in Refs.~\onlinecite{isele:15,phil:22_1,faze:23,wink:25}, as long as the bond forces are sufficiently strong, with small bond-length fluctuations such that $|{\bm r}_{i+1}-{\bm r}_{i}| \simeq r_0$.
We verified that the present choice of the active force ensures better numerical stability.
The Newton's equations of motion of the beads are integrated by the velocity-Verlet algorithm~\cite{alle:87}.

\REV{The polymer is immersed in a Brownian heat bath which is realized by adopting the Brownian Multiparticle Collision Dynamics (B-MPC) method~\cite{kiku:03,ripo:07,gomp:09}, hence, hydrodynamic interactions are absent.
This is a reasonable assumption, for example, in systems of actin filaments or microtubules on a motility assay or microorganisms that glide on a surface~\cite{scha:10,kawa:08,liu:11,keya:20}. The implemented variant of the algorithm realizes stochastic processes between a bead and the surrounding fluid volume via collisions with a single solvent particle~\cite{ripo:07}. The momentum of the virtual particle of mass $\gamma m$ is sampled from a Maxwell-Boltzmann distribution of
variance $\gamma m k_B T$ and the mean $(\gamma m \dot{\gamma} y,0)^T$ in the case of a shear flow along
the $x$-direction \REV{of the Cartesian reference frame} with the shear rate $\dot \gamma$. The collision process is performed by the stochastic rotation dynamics implementation of the MPC method~\cite{ripo:07,lamu:24}. Here, the relative velocity of a polymer bead, with respect to the
center-of-mass velocity of the bead and its related solvent particle, is randomly rotated in the $xy-$plane by an angle $\pm \alpha$.}  \REV{Further details are  presented in Appendix~\ref{app:method} and in Ref.~\onlinecite{lamu:24}.} 
Collisions occur at the time interval $\Delta t$ with $\Delta t > \Delta t_p$, where $\Delta t_p$ is the time step of the velocity-Verlet algorithm. 

The system is characterized in terms of three dimensionless numbers:
The persistence length-to-polymer-length ratio 
\begin{equation}
\frac{L_p}{L} = \frac{2 \kappa r_0}{k_B T L} ,
\end{equation}
with $L_P$ the persistence length, the P\'eclet number~\cite{isele:15,phil:22_1}
\begin{equation}
Pe = \frac{f_a N L}{k_B T} , 
\end{equation}
and the Weissenberg number 
\begin{equation}
Wi = \dot{\gamma} \tau_R  ,
\end{equation}
where $k_B T$ is the thermal energy, with $T$ the temperature and $k_B$ Boltzmann's constant.
The time $\tau_R(Pe)$ is the activity-dependent end-to-end vector relaxation time at zero shear,
which decreases as $1/Pe$ \cite{bian:18,faze:23,teje:24,pand:25.2}.
The quantity $L_p/L$ is a measure of the bending rigidity of a polymer, the 
P\'eclet number expresses
the ratio of the total active energy to the thermal energy, and 
the Weissenberg number characterizes the flow strength with respect to the characteristic relaxation time of the polymer. 

Simulations are performed with the rotation angle $\alpha=130^{\circ}$, the collision time step
$\Delta t=0.1 t_u$, with the time unit $t_u=\sqrt{m r_0^2/(k_B
T)}$, the mass of a bead $M =\gamma m$ with $\gamma=5$, and the time step $\Delta t_p=10^{-2} \Delta t$. \REV{According to Eq.~\eqref{app:friction}, this yields the ratio between the solvent friction and the bead
mass $\xi/M \approx 0.8/\Delta t \approx 8/t_u$. Hence, the inertial effects are damped out on the scale of one time unit, and do not play any role in our studies. In particular, the relaxation time $M/\xi$ is much smaller than the shortest polymer relaxation $\tau_R \approx 3\times 10^2 t_u$ considered in the current studies.}    
For the potentials, we set $\kappa_h r_0^2/(k_B T)=4 \times 10^3$, $\epsilon / (k_B T)=1$, and $r_0= \sigma$. At large values of the shear rate $\dot\gamma$ and/or of the force $f_a$ such
that $Wi Pe > 10^5$, we set  $\kappa_h r_0^2/(k_B T)=4 \times 10^4$ to avoid overstretching of the bonds under nonequilibrium conditions.  The persistence \REV{lengths } $L_p/L=0.4, \ 2$, typical of semiflexible polymers, are considered, with the number of beads $N=51$. The values of the shear rate $\dot\gamma$ are varied \REV{in} the range $1 \leq Wi \leq 10^4$. Active polymers are simulated for the P\'eclet numbers $Pe \leq 2 \times 10^4$.
At small $Pe$ ($\lesssim 10^2$), the more flexible filament with $L_p/L = 0.4$ is in the ``polymer regime'' \cite{isele:15} and \REV{behaves as a passive polymer}. When $Pe \gtrsim 10^3$, the polymer is in the ``spiral regime'', \REV{where} it is able to coil in a temporary spiral configuration whose lifetime increases with activity \cite{isele:15}. The stiffer polymer is always found in the ``polymer regime''.

The shear rates of the active polymers are chosen such to achieve values up to $Wi Pe \lesssim 10^6$ for $L_p/L=0.4$ and $Wi Pe \lesssim 5 \times 10^6$ for $L_p/L=2$, still ensuring stable simulations.

\section{Results} \label{sec:results}

Initially, the conformations and dynamics of passive and active semiflexible polymers without external flow are studied and 
the their end-to-end vector relaxation times $\tau_R$ are determined by 
B-MPC simulations.  Subsequently, the effects of shear flow on the overall polymer properties are investigated.
In all the simulations, the polymers are initialized in a straight conformation and are equilibrated up to the time $2 \tau_R(Pe=0)$. Afterwards, data are collected over a time period longer than $10 \tau_R(Pe=0)$ for the analysis.

\subsection{Conformations} \label{subsec:conform}

\begin{figure}[ht]
\begin{center}
\includegraphics*[width=.99\columnwidth,angle=0]{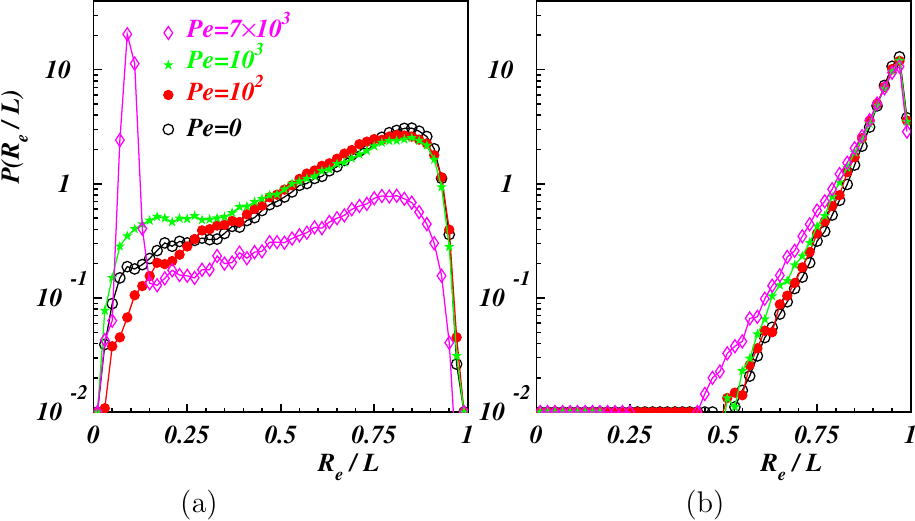}
\caption{Normalized probability distribution function of the polymer end-to-end distance $R_e=|\bm R_e|$ for (a) $L_p/L=0.4$ and (b) $L_p/L=2$, and the P\'eclet numbers 
$Pe=0 (\circ), 10^2 (\bullet), 10^3 (\star), 7 \times 10^3 (\diamondsuit)$
in absence of shear flow ($Wi=0$).
\label{fig:pdflen_wi0}
}
\end{center}
\end{figure}

The polymer conformations are characterized by the end-to-end vector ${\bm R}_e={\bm r}_N-{\bm r}_1$. Figure~\ref{fig:pdflen_wi0} presents normalized probability distribution functions (PDF) of the end-to-end distance $R_e =|\bm R_e|$ for various activity values in absence of shear ($Wi=0$). 
For both persistence lengths, there is a prevalence of conformations with large $R_e$ values.
The PDF is broad for the less stiff filament, while it becomes narrower and the peak moves toward larger
values of $R_e$ with increasing bending rigidity. Activities $Pe<10^2$ hardly affect the PDF, independent of stiffness.
However, for the more flexible polymer ($L_p/L =0.4$), larger values of activity ($Pe \gtrsim 4\times 10^3$) lead to the appearance of longer-lived spiral
and folded states~\cite{isele:15}, indicated by the sharp peak at $R_e/L \simeq 0.1$
(Fig.~\ref{fig:pdflen_wi0}(a)). The conformations of the stiffer polymers are hardly affected by activity and the PDF does not change in a significant manner, except for the largest considered activity, where PDF values at $R_e/L \approx 0.5$ are more frequent compared to the cases with $Pe \leq 10^3$ (Fig.~\ref{fig:pdflen_wi0}(b)).

\begin{figure}[b]
\begin{center}
\includegraphics*[width=.99\columnwidth]{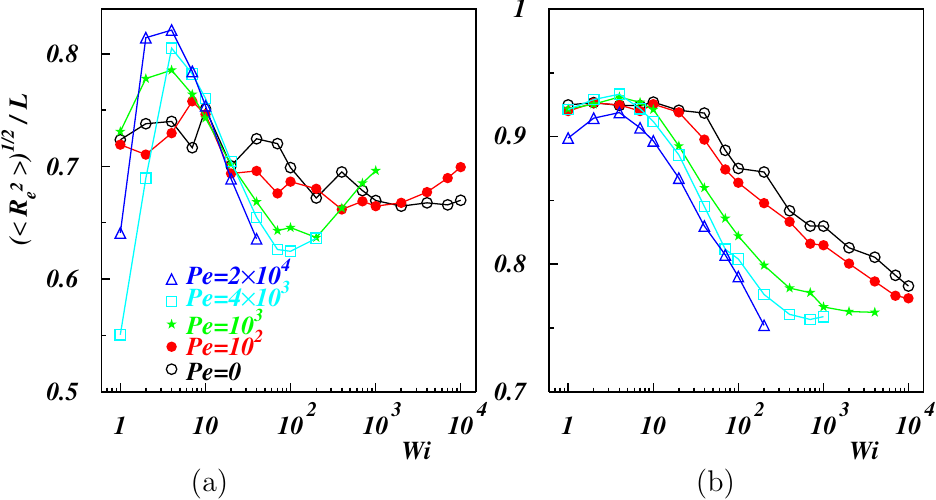}
\caption{Root mean-square end-to-end distance  
as a function of the Weissenberg number $Wi$
for (a) $L_p/L=0.4$, (b) $L_p/L=2$, and the activities
$Pe=0 (\circ), 10^2 (\bullet), 10^3 (\star), 4 \times 10^3 (\square), 2 \times 10^4 (\triangle)$. 
\label{fig:re2}
}
\end{center}
\end{figure}

\begin{figure}[ht]
\begin{center}
\includegraphics*[width=.9\columnwidth]{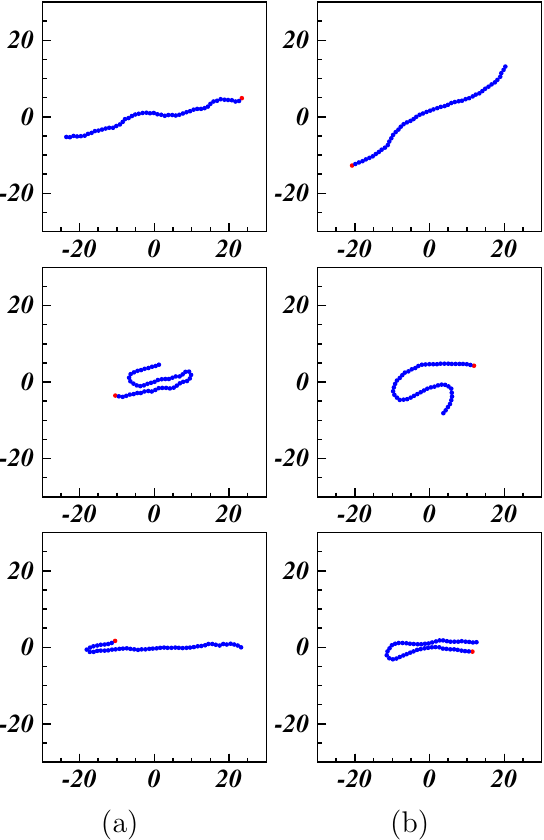}
\caption{\REV{Typical configurations of a polymer in the $xy$-plane at $Pe=10^3$
for (a) $L_p/L=0.4$, (b) $L_p/L= 2$, and
$Wi=4, 70, 10^3$ (from top to bottom). The red bead indicates the leading end.}
\label{fig:conf_pe1000}
}
\end{center}
\end{figure}

The root mean-square end-to-end distance $\lla R_e^2\rra^{1/2}/L$ of the polymer is displayed in Fig.~\ref{fig:re2} as a function of the Weissenberg number. 
The values of $\lla R_e^2\rra^{1/2}/L$ of the more flexible polymer are always smaller than to those of the stiffer polymer.
Figure \ref{fig:re2}(a) shows that the root mean-square end-to-end distance is \REV{nearly constant} at low shear rates 
in the passive case, since the polymer is initially only aligned by the shear along the flow direction. \REV{Beyond a certain $Wi$ value, 
it decreases due to shear-induced deformations, i.e., by the appearance of folds and hairpin-like configurations (see snapshots in Fig.~\ref{fig:conf_pe1000}), and, finally, it approaches a constant value for very  high  values of $Wi$.}

In presence of activity, particularly at large Pe, shear enhances the stretching of the more flexible polymer \REV{at Weissenberg numbers $Wi \gtrsim 1$}, and $\lla R_e^2\rra^{1/2}/L$ assumes a maximum at $Wi= Wi_M \approx 4$. This maximum increases with increasing activity. \REV{Simultaneously}, at intermediate values of
the Weissenberg number, $\lla R_e^2\rra^{1/2}/L$ decreases with increasing shear rate, assumes a minimum at $Wi= Wi_m \approx 70$, and, finally,
grows again for very high values of $Wi$.   

\REV{The strong increase at small $Wi$ is related to the more compact structures of the active polymers at larger $Pe$ \REV{in absence of shear, specifically the formation of spiral structures for $Pe \gtrsim 10^3$ as illustrated in Ref.~\onlinecite{isele:15}.} 
These compact structures are unfolded by shear. Even more, activity supports stretching of a polymer over this scale of Weissenberg numbers. At larger shear rates ($Wi>Wi_M$), the polymer tumbles and shrinks. It performs contraction and stretching cycles while rotating, and
assumes hairpin and S-shaped conformations \cite{smit:99,ledu:99,schr:05,teix:05,ryde:06,gera:06,wink:06,wink:10,huan:11,dala:12}
(cf. Fig.~\ref{fig:conf_pe1000} and the multimedia file movie2.mp4 available online 
for $L_p/L=0.4$, $Pe=10^3$, $Wi=70$).}
At very high values of the Weissenberg number, the tumbling dynamics is faster and 
the polymer bends completely, enhanced by activity, while sliding over itself
(see the multimedia file movie3.mp4 available online for $L_p/L=0.4$, $Pe=10^3$, $Wi=10^3$).

\begin{figure}[ht]
\begin{center}
\includegraphics*[width=.8\columnwidth]{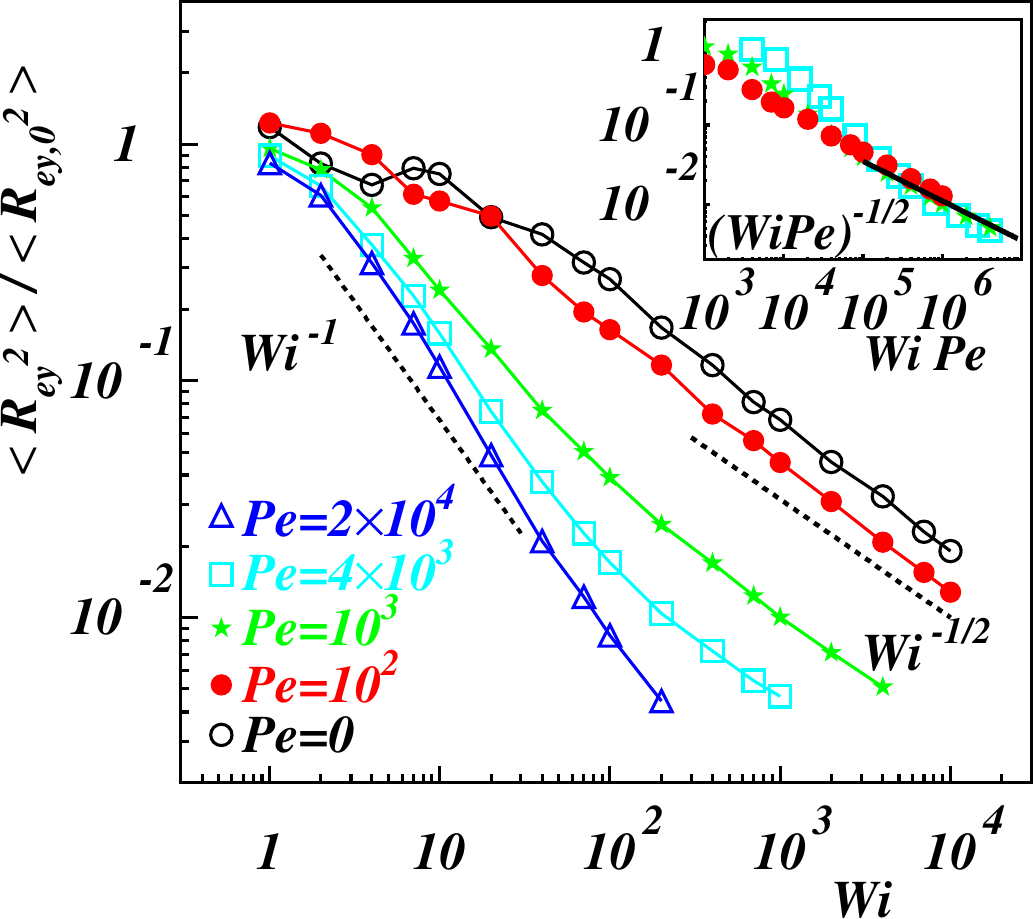}
\caption{Mean-square end-to-end distance along the gradient direction ($y$-axis)
as a function of the Weissenberg number $Wi$ for $L_p/L=2$ and
$Pe=0 (\circ), 10^2 (\bullet), 10^3 (\star),
4 \times 10^3 (\square), 2 \times 10^4 (\triangle)$. 
$\langle R^2_{ey,0} \rangle$ is the mean-square end-to-end distance in absence of shear ($Wi=0$).  Dashed 
lines indicate the slopes $-1/2$ and $-1$.
Inset: Mean-square end-to-end distance
as a function of $Wi Pe$ for $Pe>0$. The solid line indicates the dependence $(WiPe)^{-1/2}$.} 
\label{fig:ry2}

\end{center}
\end{figure}

The stiffer polymer (Fig.~\ref{fig:re2}(b)) is mainly aligned along the flow direction at low values of $Wi \lesssim 10$,\cite{wink:10} and 
$\lla R_e^2\rra^{1/2}/L$ only weakly depends on activity for $Pe \lesssim 10^2$. \REV{As for the more flexible polymer, with increasing $Pe > 10^3$, the polymer assumes slightly more compact conformations, which are ''unfolded'' by shear and lead to a small maximum.}  
For $Wi>10$, the root-mean-square end-to-end distance decreases, with a substantially stronger drop at large activities, $Pe \gtrsim 10^2$, in an intermediate shear-rate regime, again due to tumbling. The reason of this quantitative difference between an active and a passive polymer will become more evident, when considering the mean-square end-to-end distance in the gradient direction presented in Fig.~\ref{fig:ry2}.
At large, $Pe$-dependent Weissenberg numbers, $\lla R_e^2\rra^{1/2}/L$ assumes a shear-rate independent value. A similar qualitative behavior is displayed by the radius of gyration \REV{(not shown)}.


\begin{figure}[ht]
\begin{center}
\includegraphics*[width=.99\columnwidth]{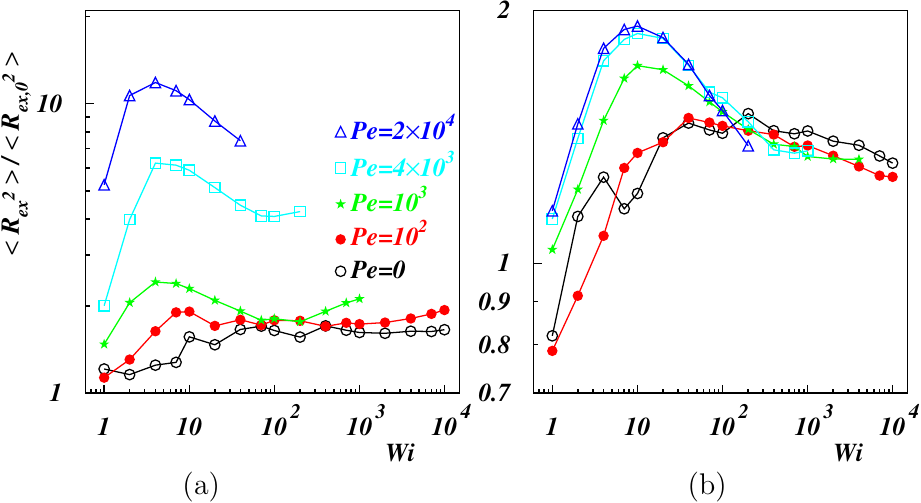}
\caption{Mean-square end-to-end distance 
along the flow direction ($x$-axis)
as a function of the Weissenberg number $Wi$
for (a) $L_p/L=0.4$, (b) $L_p/L= 2$, and
$Pe=0 (\circ), 10^2 (\bullet), 10^3 (\star),
4 \times 10^3 (\square), 2 \times 10^4 (\triangle)$.
$\langle R^2_{ex,0} \rangle$ is the mean-square end-to-end distance in absence of shear ($Wi=0$).
\label{fig:rx2}
}
\end{center}
\end{figure}

The ratio of the mean-square end-to-end distances, $\langle R_{ex}^2\rangle/\langle R_{ex,0}^2\rangle$, along the flow direction exhibits similar features, as shown
in Fig.~\ref{fig:rx2}. Clearly, shear enhances the stretching of a polymer with respect to the unsheared case, where the relative stretching is much larger for the more flexible polymer, specifically for large activities. This is related to the more compact polymer conformation in absence of shear and its ''unfolding'' and stretching with increasing shear rate. 

The mean-square end-to-end distance in the gradient direction, $\langle R_{ey}^2 \rangle$, decreases with increasing shear rate, as a consequence of the polymer alignment and stretching along the flow direction, independent of the polymer stiffness \cite{wink:10,huan:10}. Figure~\ref{fig:ry2} illustrates the shear-rate dependence of $\lla R_{ey}^2\rra$ for the stiffer polymer.
The decay in the passive case is given by the power law
$\langle R_{ey}^2 \rangle \sim Wi^{\nu}$ with $\nu \approx -1/2$  in the large $Wi$ limit, consistent with previous studies~\cite{lamu:12,huan:10}. Remarkably, the decay is faster in presence of activity, and a transient scaling regime at intermediate values of $Wi$ with the power-law exponent $\nu \approx -1$ is assumed in the large $Pe$ limit. This behavior reflects the fact that shear facilitates the 
stretching of an active polymer by suppressing its conformational fluctuations and simultaneously aligns it in the flow direction. 
This holds for both values of the persistence length. In the asymptomatic limit $Wi \to \infty$, shear dominates over activity, and the polymer behaves as a passive polymer \cite{pand:25,pand:25.2}. The crossover to the passive regime depends on the P\'eclet number. \REV{The inset of Fig.~\ref{fig:ry2} illustrates the asymptotic dependence of $\langle R_{ey}^2 \rangle$ on $Wi Pe$ for $Pe \leq 4 \times 10^3$, as larger Weissenberg numbers are required to reach the passive limit for $Pe=2\times 10^4$. Since the relaxation time $\tau_R \sim 1/Pe$ and $Wi = \dot \gamma \tau_R$, the product is independent of the activity in this limit.}   

A stronger shrinkage of the polymer along the gradient direction in presence of activity, compared to the passive polymers under shear, has also been obtained for active Brownian polymers \cite{mart:18,pand:23}, polar linear flexible polymers in three dimensions \cite{pand:25,pand:25.2}, as well as ring polymers \cite{wink:24}. In case of the flexible active Brownian polymer, the exponent $\nu \approx -3/4$ is obtained in presence of excluded-volume interactions \cite{pand:23}, whereas theory for stiff phantom polymers suggests the exponent $\nu \approx -1$ \cite{mart:18}. Simulations and theory of active polar linear polymers yield the value $\nu \approx -4/3$ \cite{pand:25,pand:25.2}. The exponents obtained for flexible active polar polymers in three dimensions differ from that obtained of the here considered semiflexible polymers in two dimensions. We attribute this difference to stiffness of our considered polymers. Even more, theory yields the relation $Wi/\sqrt{Pe}$ for the crossover from the activity-dependent to the passive polymer regime  for flexible phantom polymers in three dimensions. Thus, the characteristic relaxation time for the crossover depends on activity, in contrast to the stiffer active polar polymers in two dimensions. It remains to be shown, to which extent excluded-volume interactions are relevant for the crossover, since these interactions matter for the shrinkage in the gradient direction of passive polymers under shear~\cite{huan:10}.      

\begin{figure}[ht]
\begin{center}
\includegraphics*[width=.99\columnwidth]{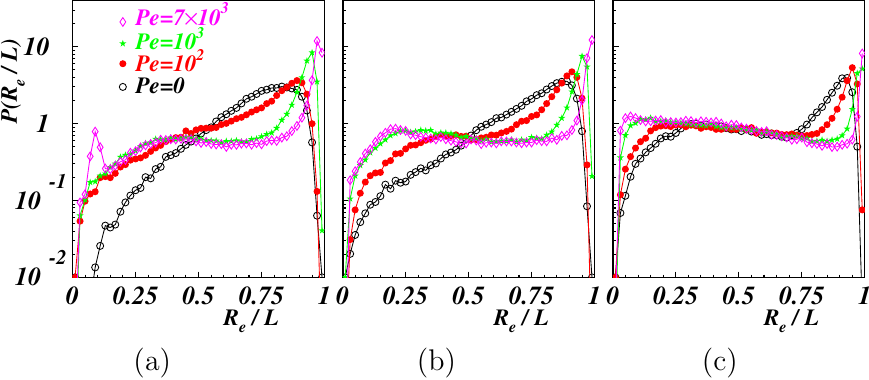}
\caption{Normalized probability distribution function
of the polymer root-mean-square end-to-end distance
for $L_p/L=0.4$ and  (a) $Wi=4$, (b) $Wi=10$, (c) $Wi=10^2$, and 
$Pe=0 (\circ), 10^2 (\bullet), 10^3 (\star), 7 \times 10^3 (\diamondsuit)$.
\label{fig:pdflen_04}
}
\end{center}
\end{figure}

\begin{figure}[ht]
\begin{center}
\includegraphics*[width=.99\columnwidth]{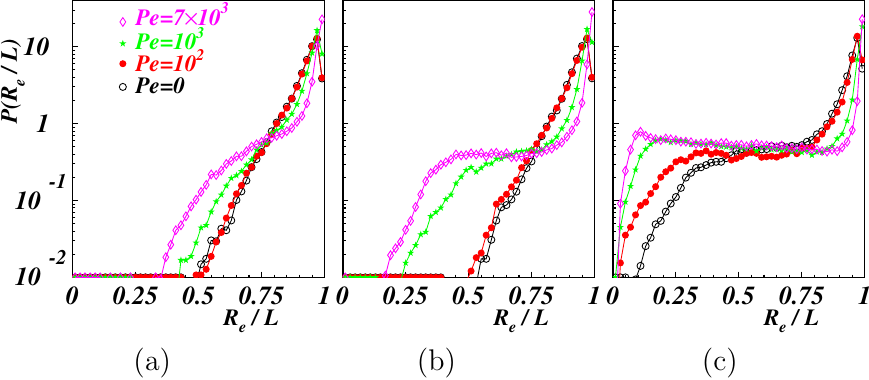}
\caption{Normalized probability distribution function
of the polymer root-mean-square end-to-end distance
for $L_p/L=2$ and  (a) $Wi=4$, (b) $Wi=10$, (c) $Wi=10^2$, and 
$Pe=0 (\circ), 10^2 (\bullet), 10^3 (\star), 7 \times 10^3 (\diamondsuit)$.
\label{fig:pdflen_2}
}
\end{center}
\end{figure}

The normalized probability distribution functions of the polymer root-mean-square
end-to-end distance in presence of shear are presented in 
Figs.~\ref{fig:pdflen_04} and \ref{fig:pdflen_2} for the two stiffnesses. 
Each figure displays the PDFs at 
different values of the Weissenberg number and various $Pe$ values.
For the more flexible polymer, $L_p/L=0.4$, shearing leads to stretching of the polymer, which increases with increasing activity. \REV{Shear suppresses conformational changes for the active polymer, and stretched conformations dominate the distribution function. An increasing activity leads to a narrowing and (small) shift of the main peak toward larger $R_e$ values compared to the passive case.} 
Simultaneously, the probability $P(R_e/L)$ of smaller $R_e$ values increases at a given shear rate, which corresponds to an enhanced polymer flexibility. \REV{ Similarly, at a given $Pe$, an increasing Weissenberg number implies larger polymer deformations and the presence of more compact polymer conformations, as the peak at lower $Pe$ shifts to smaller vales.}  Yet, stretched conformations dominate the distribution function.  

The observed behavior points to the following main physical features of the present systems:
\begin{itemize}
\item At low values of $Wi \lesssim 10$, the semiflexible polymers are aligned by shear along the flow direction. Simultaneously, activity-induced compact conformations, specifically temporary spiral structures~\cite{isele:15}, are resolved and the polymer is stretched. This effect is most pronounced for flexible polymers
and extends up to the maxima of $\lla R_{e}^2\rra$ and $\lla R_{ex}^2\rra$; 
\item At intermediate values of $Wi$, polymers are prone to tumble, as  will be shown in the following. This promotes a folding of a polymer, corresponding to the minima of $\lla R_{e}^2\rra$ and $\lla R_{ex}^2\rra$, and to the emergent second peak in the PDF's of $R_e$ at small $R_e$;
\item At very high values of $Wi$, shear dominates over activity and the polymers behave as passive ones.
\end{itemize}

\begin{figure}[t]
\begin{center}
\includegraphics*[width=.99\columnwidth]{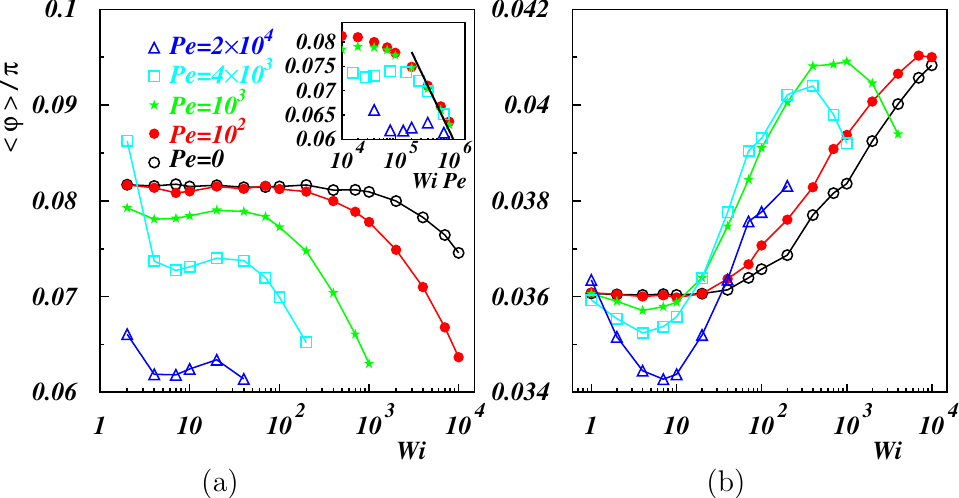}
\caption{Average bond angle $\lla \varphi \rra$
as a function of the Weissenberg number $Wi$
for (a) $L_p/L=0.4$, (b) $L_p/L=2$, and various P\'eclet numbers (legend). 
Inset in (a): Average bond angle as a function
of $Wi Pe$ for $Pe>0$. The solid line is a guide to the eye.
\label{fig:angle}
}
\end{center}
\end{figure}

More detailed insight into the polymer conformational properties is gained by the 
average bond angle $\langle \varphi \rangle =  \sum_{i=1}^{N-2} \langle \varphi_i \rangle  /(N-2)$, where $\varphi_i$ is the angle  between subsequent bonds along the polymer contour. Figure~\ref{fig:angle} displays simulation results as a function of the Weissenberg number.
In the flexible case (Fig.~\ref{fig:angle}(a)) and $Pe \leq 10^2$, $\langle \varphi \rangle$ exhibits a plateau and decreases for $Wi>10^3$ as a \REV{consequence of the shear-induced stretching of parts of the polymer (Fig.~\ref{fig:conf_pe1000})~\cite{lamu:12}.} \REV{The average angle of the stiffer polymer increases with increasing shear for P\'eclet numbers $Pe \gtrsim 10^2$ (Fig.~\ref{fig:angle}(b)). } 
For larger activities, $Pe \gtrsim 10^3$, a nonmonotonic behavior appears. At small $Wi \lesssim 3$, folded, spiral-type configurations occur, with large bond angles.  \REV{\REV{Larger Weissenberg numbers, $1< Wi <10 $,  enhance the activity-induced polymer stretching (cf.  Figs.~\ref{fig:re2}, \ref{fig:rx2})}, with a corresponding decrease in $\langle \varphi \rangle$. For larger Weissenberg numbers, shear gradually dominates over activity, which implies a reduction of the polymer stretching along the flow direction and an associated increase of $\langle \varphi \rangle$. Finally, above a stiffness- and activity-dependent Weissenberg number, shear dominates over activity and a polymer behaves as a passive one. Here, shear stretches and folds a polymer, which reduces  $\langle \varphi \rangle$, and a maximum appears, specifically for the stiffer polymer (cf. Fig.~\ref{fig:angle}(b) and the multimedia files movie1.mp4, movie2.mp4, and movie3.mp4 available online).}  The inset of Fig.~\ref{fig:angle}(a) suggest a weak logarithmic dependence of the average bond angle on $Wi Pe$ for large $Wi$.

As expected, the mean bond angles of the stiffer polymer are smaller than those of the more flexible polymer (Fig.~\ref{fig:angle}(b)). 
At the same time, fluctuations are limited to approximately $10\%$ for the considered range of the Weissenberg numbers.
The increase of the mean bond angle for very large shear rates of the passive polymer results from the tumbling motion and the emerging  U-shaped conformations \cite{lamu:12}. Activity causes a nonmonotonic behavior, similar to the more flexible filament, although with the minima and maxima shifted to larger values of $Wi$.

\subsection{Alignment} \label{subsec:alignment}

\begin{figure}[t]
\begin{center}
\includegraphics*[width=.99\columnwidth]{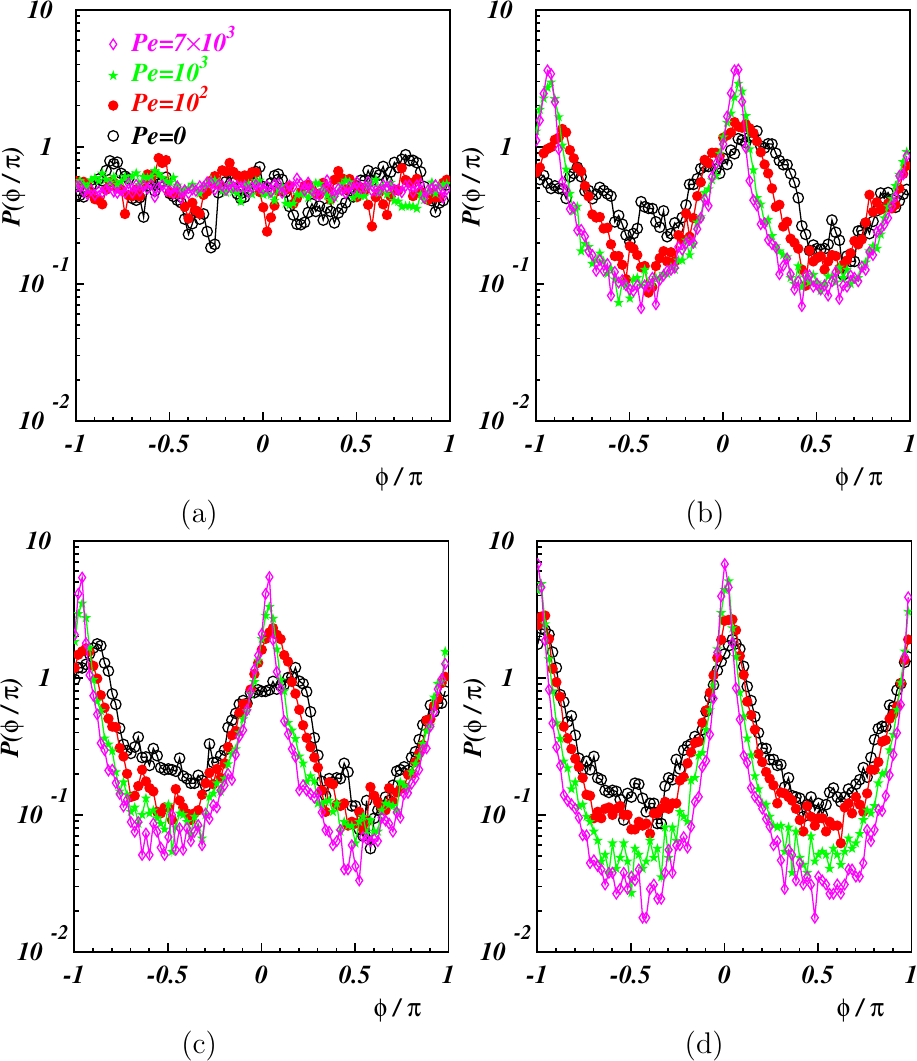}
\caption{Normalized probability distribution function
of the angle $\phi$ between the end-to-end vector and the flow direction
for $L_p/L=0.4$ and
(a) $Wi=0$, (b) $Wi=4$, (c) $Wi=10$, and (d) $Wi=10^2$. The P\'eclet numbers are
$Pe=0 (\circ), 10^2 (\bullet), 10^3 (\star), 7 \times 10^3 (\diamondsuit)$.
\label{fig:pdfang_04}
}
\end{center}
\end{figure}

As already emphasized, sheared active polymers are deformed and preferentially aligned along the flow direction.
The alignment is characterized by the probability distribution function $P(\phi)$ of the angle $\phi$ between the end-to-end vector ${\bm R}_e$ and the flow direction ($x$-axis). Results are depicted in Fig.~\ref{fig:pdfang_04} for $L_p/L=0.4$.
In absence of shear, $P(\phi)$ is uniform and the orientation of $\bm R_e$ is isotropic. The large fluctuations of the distribution function for $Pe=0$ are a consequence of a slow sampling of the angles. These fluctuations are progressively reduced  with increasing $Pe$ by a  more rapid exploration of all possible directions. In presence of shear, two peaks of comparable height emerge in the vicinity of $\phi_m \approx 0^+$ and $\phi_m \approx-\pi^+$. These peaks indicate the preferential alignment of a polymer with equal probability \REV{along the flow direction with the trailing end located in the positive or negative gradient direction, respectively.}
At the same time, the orientation in the gradient direction is disfavored, reflected by the minima at $\phi \approx \pm \pi/2$.  With increasing shear rate, the maxima $\phi_m$ shift to smaller values and the peaks become narrower, indicating a stronger alignment along the flow direction. 

\begin{figure}[t]
\begin{center}
\includegraphics*[width=.8\columnwidth]{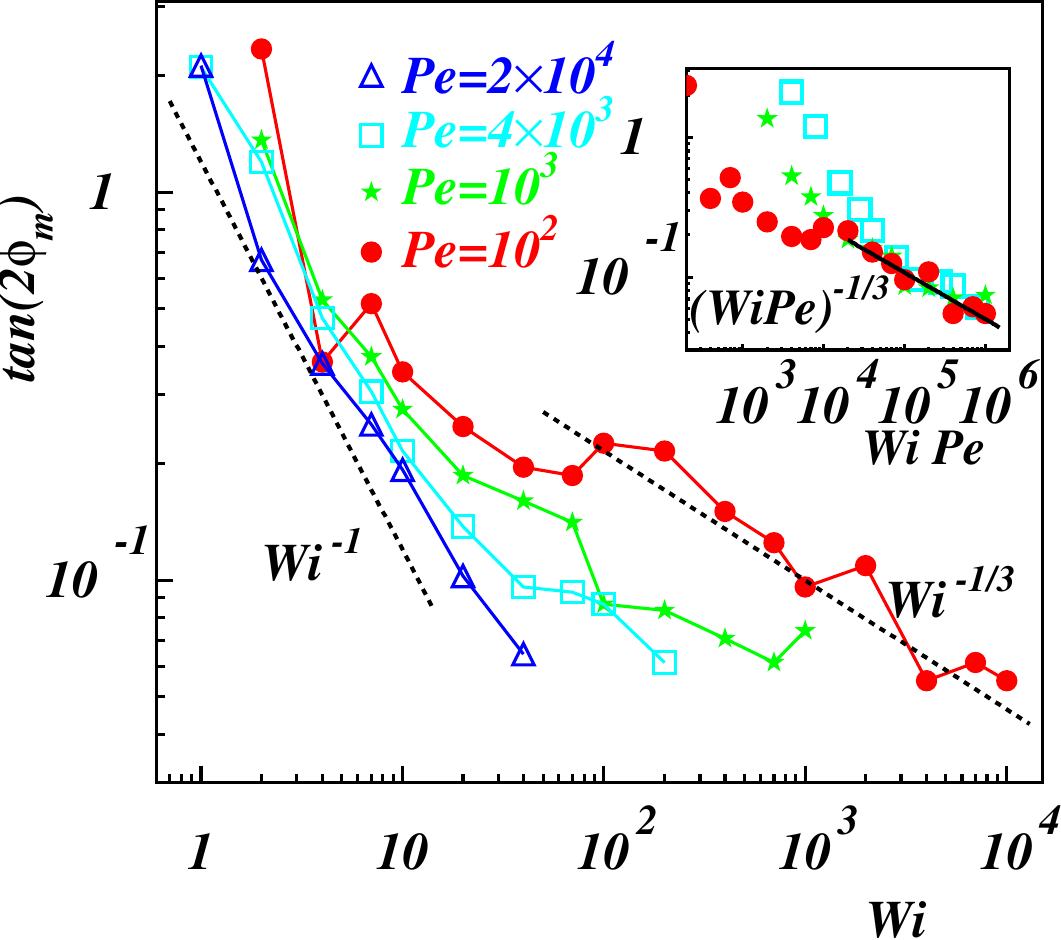}
\caption{Angle $\phi_m$ of the maxima of the distribution functions 
$P(\phi)$ as a function of the Weissenberg number for the persistence
length $L_p /L = 0.4$ and the P\'eclet numbers
$Pe=10^2 (\bullet), 10^3 (\star),
4 \times 10^3 (\square), 2 \times 10^4 (\triangle)$. 
The dashed lines indicate the  slopes $-1$ and $-1/3$. 
Inset: Angle $\phi_m$ as a function of $Wi Pe$ for $Pe>0$.
The solid indicates the dependence $(WiPe)^{-1/3}$.
\label{fig:tan}
}
\end{center}
\end{figure}

The values $\tan(2 \phi_m)$ of the angle $\phi_m$ are displayed in Fig.~\ref{fig:tan} for the more flexible polymer ($L/L_p=0.4$).
For the smallest values of the activity ($Pe=10^2$) and $Wi \gtrsim 10$, we find the dependence $\tan(2 \phi_m) \sim Wi^{-1/3}$ similar to that of a passive polymer at large shear rates~\cite{lamu:12,wink:10,huan:10}. With increasing
$Pe$, an extended additional regime appears, where $\tan(2 \phi_m)$ exhibits the decay $\tan(2 \phi_m) \sim Wi^{-1}$ over a $Pe$-dependent shear-rate regime, which increases with increasing $Pe$, and crosses over to the passive polymer behavior for larger $Wi$. Passive polymers exhibit a similar decay at small Weissenberg numbers due to their alignment along the flow direction~\cite{huan:10,wink:10}. However, activity extends that regime toward larger Weissenberg numbers. A similar behavior is obtained for flexible polymers in three dimensions by simulations~\cite{pand:25} and analytical theory~\cite{pand:25.2}.

The inset of Fig.~\ref{fig:tan} illustrates the approach to the passive limit at large Weissenberg numbers. In particular, it reveals the scaling relation 
\begin{equation}
\tan(2 \phi_m) \sim (Wi Pe)^{-1/3}
\end{equation}
in terms of the P\'eclet and the Weissenberg number \REV{for $Pe \leq 4\times 10^4$. Larger $Pe$ require larger Weissenberg numbers than those applied to reach the passive limit.} Since $Wi \sim 1/Pe$, alignment is independent of activity and equal to that of passive polymers in the asymptotic limit of large $Wi$. The same conclusion has been drawn for the shear and activity dependence of the mean-square end-to-end distance along the shear direction in Sec.~\ref{subsec:conform}. 

The analogous trend is followed for the stiffer filament although the values of the angle $\phi_m$ are slightly larger compared to the more flexible polymer, consistent with theoretical predictions of passive polymers \cite{wink:10}.

\subsection{Tumbling dynamics} \label{subsec:dynamics}

\begin{figure}[t]
\begin{center}
\includegraphics*[width=.8\columnwidth]{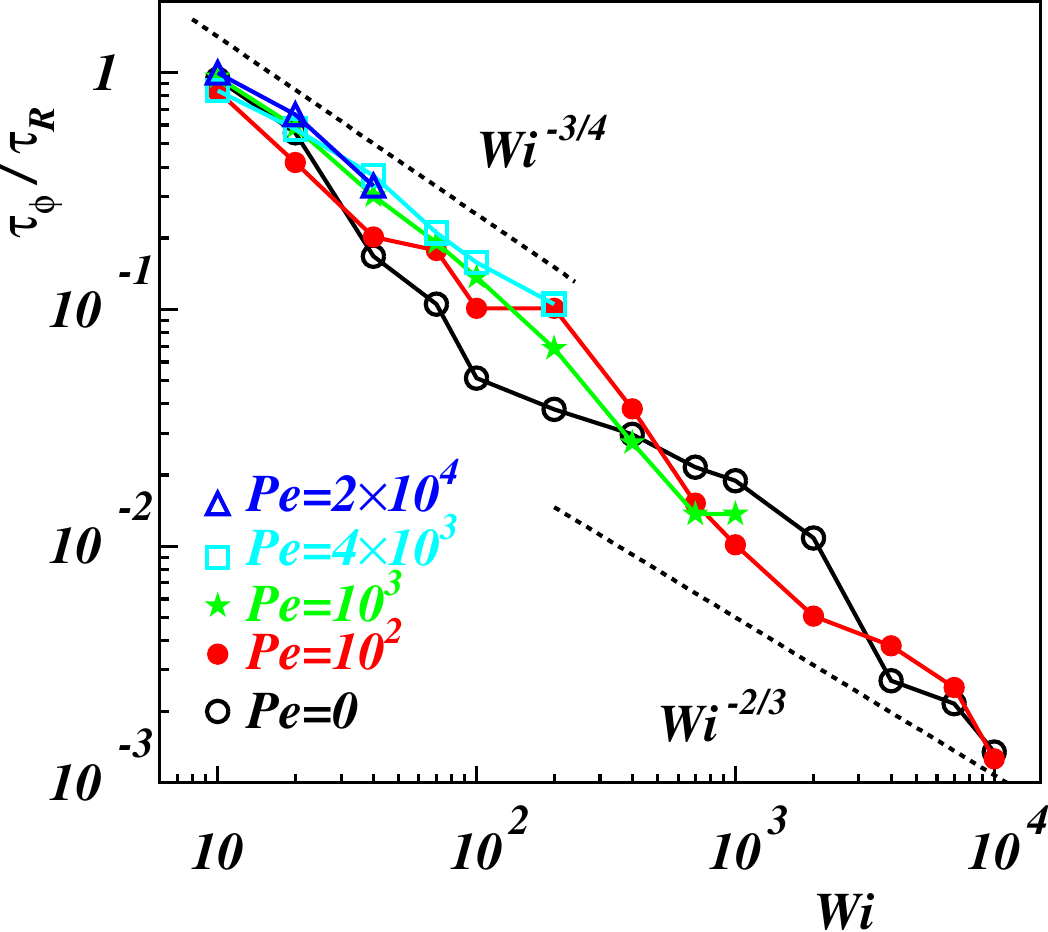}
\caption{
Normalized tumbling time $\tau_{\phi}/\tau_R$ as a function of the Weissenberg number for the persistence
length $L_p /L = 0.4$ and the P\'eclet numbers $Pe=0 (\circ), 10^2 (\bullet), 10^3 (\star),
4 \times 10^3 (\square), 2 \times 10^4 (\triangle)$. 
The dashed lines indicate the slopes $-3/4$ and $-2/3$.
\label{fig:tumbl}
}
\end{center}
\end{figure}

A characteristic time for the cyclic tumbling dynamics of the polymers can be extracted from the distribution function $P(t)$ of time intervals $t$, between subsequent sign changes of the end-to-end vector component $R_{ex}(t)$. At long times, this distribution function exhibits the dependence $P(t) \sim \exp(-t/\tau_{\phi})$, so that the characteristic time, the tumbling time $\tau_{\phi}$, can be extracted~\cite{wink:06.1,wink:24}. The tumbling times $\tau_{\phi}$ determined from $P(t)$ for the more flexible polymer are displayed in Fig.~\ref{fig:tumbl}. Data collapse for all activity values is achieved by
rescaling $\tau_{\phi}$ by the respective relaxation times $\tau_R$ of the unsheared case. Two shear-rate dependent regimes can be identified.
At intermediate values of shear, the dependence $\tau_{\phi}/\tau_R \sim Wi^{-3/4}$ is observed. At larger shear rates,
the decay $\tau_{\phi}/\tau_R \sim Wi^{-2/3}$ is found, which is typical 
of passive semiflexible polymers \cite{schr:05,puli:05,wink:06,huan:10,wink:10,pinc:23} and Brownian rods \cite{munk:06,koba:10}. Depending on the applied polymer model, various simulation studies suggest a crossover to the power-law $Wi^{-3/4}$ at higher shear rates for semiflexible polymers \cite{munk:06,dala:12}. In contrast, the current simulations of active polar polymers show this dependence for small shear rates, which points toward the activity dependence of this regime. This is consistent with the conformational changes, e.g., of $\langle R_{ey}^2 \rangle$ in Fig.~\ref{fig:ry2}, over the same Weissenberg number interval. 

Simulations of three-dimensional flexible polar polymers yield the same passive limit $\tau_{\phi} \sim Wi^{-2/3}$ at large shear rates. However, at small Weissenberg numbers, the power-law $\tau_{\phi} \sim Wi^{-1/3}$ is obtained. This pronounced difference in the exponent could be related to the substantial conformational changes of the semiflexible polymer along the flow direction (cf. Fig.~\ref{fig:rx2}(b)) in the range $10< Wi <200$.  

\begin{figure}[t]
\begin{center}
\includegraphics*[width=.8\columnwidth]{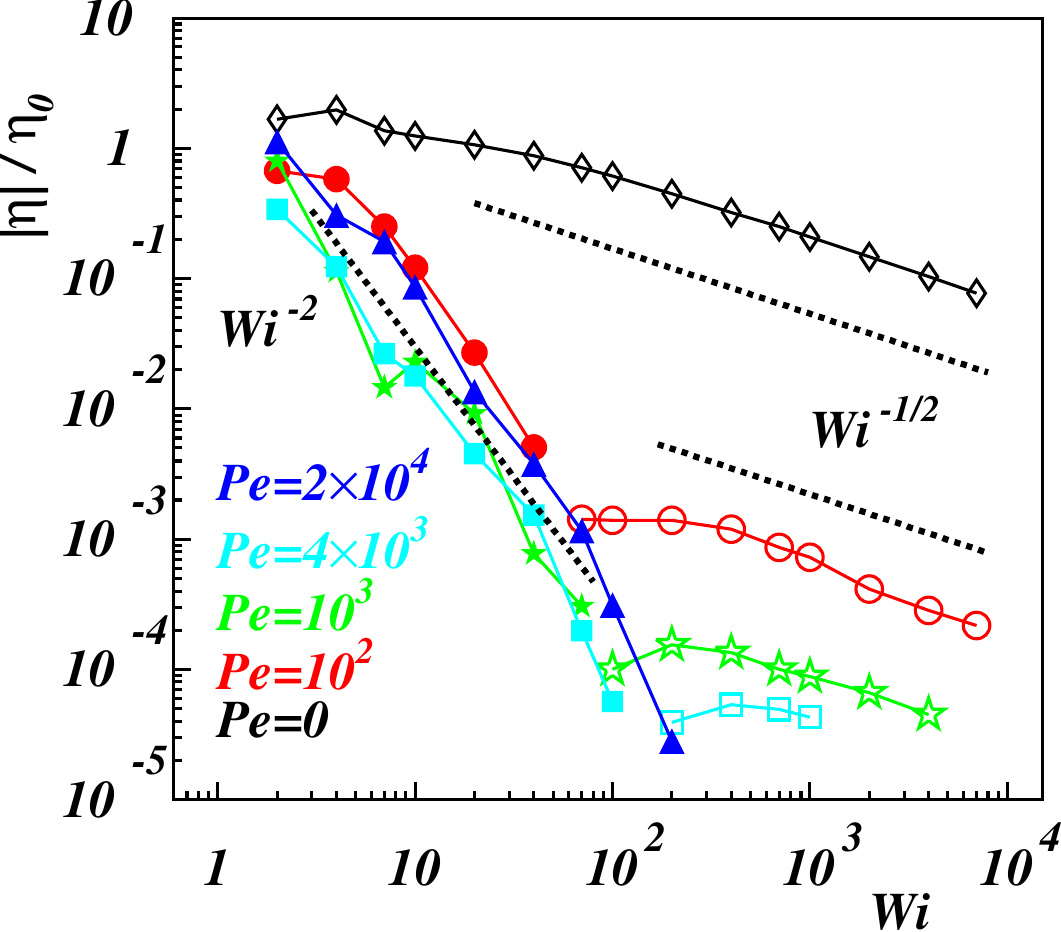}
\caption{
\REV{Magnitude of the normalized polymer contribution to the viscosity, $\eta/\eta_0$, as a function of the Weissenberg number for the persistence length
$L_p /L = 2$ and the P\'eclet numbers $Pe=0 (\Diamond), 10^2 ({\Large\bullet}), 10^3 (\star),
4 \times 10^3 (\blacksquare), 2 \times 10^4 (\blacktriangle)$; $\eta_0$ is the $Pe$-dependent zero-shear viscosity. Filled and open symbols correspond to the negative and positive
viscosity values, respectively.
The dashed lines indicate the slopes $-2$ and $-1/2$.}
\label{fig:viscosity}
}
\end{center}
\end{figure}

\subsection{Viscosity} \label{sec:viscosity}

\REV{To elucidate the influence of stiffness and activity on the rheological properties of the polymers, we determine their contribution to the shear viscosity. The viscosity, $\eta$, itself follows from the virial expression of the stress tensor via $\eta=\sigma_{xy}/\dot \gamma$, where ~\cite{wink:10,pand:25.2,bird:87} 
\begin{equation}
    \sigma_{xy} = - \frac{1}{A} \sum_{i=1}^N (F_{ix}^{in} + F_{ix}^a) r_{yi} ,
\end{equation}
with the intramolecular forces $\bm F_i^{in}$,  following from Eqs.~\eqref{bond}-\eqref{rep_pot}, 
the active force of Eq.~\eqref{force}, and $A$ being the system area.}

\REV{Figure~\ref{fig:viscosity} displays the simulation results for $L_p/L=2$ and various activities. In the passive limit, the viscosity decreases approximately with the exponent $-1/2$ with increasing shear, as observed and predicted previously~\cite{aust:99,hsie:04,huan:10,schr:05.1,pand:23,pand:25}. Remarkably, above a certain P\'eclet number, the viscosity is negative over an activity-dependent range of Weissenberg numbers, as indicted in Fig.~\ref{fig:viscosity} for $Pe\geq 10^2$. As already mentioned, with increasing shear rate, shear dominates over activity, and a polymer behaves like a passive polymer. The viscosity is then positive and decreases approximately with the exponent $-1/2$ with increasing shear rate. In the low Weissenberg-number regime, the negative viscosity increases roughly quadratically with increasing shear rate. Moreover, it is nearly independent---or only weakly dependent---on activity. }      

\REV{The negative viscosity is a consequence of the stiffness, because the analytical calculation for flexible polar polymers yield a positive viscosity~\cite{pand:25.2}. An analytical calculation based on the Gaussian dumbbell model~\cite{wink:16} with polar forces~\cite{suma:14.1} yields two contributions to the viscosity, a positive contribution by the bond forces and a negative one due to the active forces. The latter shows a quadratic dependence on the active force $f_a$, its relaxation time $\tau$, and the inverse Weissenberg number, i.e., $\eta \sim - (f_a \tau /Wi)^2$. This explains qualitatively the emergent behavior of the semiflexible polymers.}
 
\section{Conclusions} \label{sec:conclusions}

We have presented simulation results for the structure, dynamics, and rheology 
of tangentially driven semiflexible polymers, confined to two dimensions, under shear flow.
Our investigations reveal a severe impact of polar activity on the properties of sheared semiflexible filaments. Qualitatively, similar rheological properties are observed as for passive polymers, however, activity implies substantial quantitative differences. 

In the absence of shear, a large activity and a low bending rigidity favor the temporal formation of spiral-like structures. 
In the opposite limit of rather rigid filaments, self-propulsion induces a motion of the polymer along its own contour, superimposed
with thermal noise \cite{isele:15}, as observed in the PDF of the end-to-end vector (Fig.~\ref{fig:pdflen_wi0}).
Shear is effective in altering the phenomenology of conformations. For Weissenberg numbers 
$Wi \lesssim 10$, the applied flow disentangles spiral structures and stretches the polymer along the flow direction. This is reflected in the mean-square end-to-end distance along the flow direction, which increases substantially, in particular, for the more flexible polymer with $L_p/L=0.4$. For larger Weissenberg numbers, the polymer performs tumbling motion, where it folds onto itself, thus, reducing its average extension. 
The stretching along the flow direction is inevitably associated with a shrinkage in the gradient direction. Here, the  dependence Weissenberg-number dependence $\lla R^2_{ey} \rra \approx Wi^{-1}$ is obtained in the large $Pe$ limit and $Wi \gtrsim 1$. This decay is faster than that of passive polymers, where the power-law exponent is $\approx -1/2$. \cite{huan:10,wink:10,lamu:12}
The exponent $-1$ seems to be characteristic for the present semiflexible polymers, as an exponent $-4/3$ is obtained for flexible active polar linear polymers 
in three dimensions by simulations and analytical studies~\cite{pand:25,pand:25.2}. At very high values of $Wi$, shear dominates over activity effects and the polymer behaves as in the passive case. The same crossover has been found for flexible polar polymers~\cite{pand:25,pand:25.2}.

The alignment along the flow direction has been scrutinized by evaluating the PDF's of the 
angle between the end-to-end vector and the flow direction. It appears that the most 
probable orientation angle $\phi_m$ diminishes with increasing shear rate and propulsion force.
For small values of activity ($Pe \lesssim 10^2$), the known dependence $\tan(2 \phi_m) \sim Wi^{-1/3}$ of passive polymers for large shear rates 
 is recovered~\cite{wink:10,huan:10,lamu:12}. When $Pe \gtrsim 10^3$, activity implies a stronger flow dependence, where  approximately
$\tan(2 \phi_m) \sim Wi^{-1}$ over an intermediate $Pe$-dependent shear-rate regime. The passive polymer behavior is recovered at large values 
of the Weissenberg. As pointed out in Sec.~\ref{subsec:alignment}, passive polymers
exhibit a similar decay at small Weissenberg numbers due to their alignment along the flow direction~\cite{huan:10,wink:10}. Yet, activity extends the regime toward larger Weissenberg numbers. A similar behavior is obtained for flexible polymers in three dimensions by simulations and analytical theory.

We have also shown that self-propelled semiflexible polymers exhibit tumbling motion. 
The tumbling time $\tau_{\phi}$ show the dependence $\tau_{\phi}/\tau_R \sim Wi^{-3/4}$ at intermediate
values of the Weissenberg number---$10<Wi < 10^2$ in the present case.  At vary large values of the shear rate, 
the decay $\tau_{\phi}/\tau_R \sim Wi^{-2/3}$, characteristic for passive semiflexible polymers
\cite{schr:05,puli:05,wink:06,huan:10,wink:10,pinc:23}, is recovered. The initial $Wi^{-3/4}$ dependence seems to be specific for semiflexible polymers, because simulations of three-dimensional flexible polar polymers yield $\tau_{\phi} \sim Wi^{-1/3}$.~\cite{pand:25} We attribute the  difference in the exponent to the substantial conformational changes of semiflexible polymers along the flow direction in the range $10< Wi <200$ (Fig.~\ref{fig:rx2}(b)). 

\REV{Calculations of the shear viscosity yield negative viscosity values at small Weissenberg numbers. Thereby, the Weissenberg-number regime depends on the P\'eclet number, and is limited by the crossover to passive polymer behavior. To gain insight into the observed effect, we approximate the semiflexible polymer by a dumbbell. An analytical calculation based on an active polar Gaussian dumbbell model confirms the observed negative viscosity and its dependence on the Weissenberg number, i.e., $\eta \sim -1/Wi^2$. Remarkably, the negative viscosity appears in a dry active system. Negative viscosities are typically predicted for wet active systems with pusher-type propulsion, with their flow field interfering with the external flow field~\cite{hatw:04,sain:10,marc:13}. Further studies are necessary to gain a deeper understanding of the observed viscosity properties. Specifically, the critical P\'eclet number for the crossover from the negative to the positive viscosity needs to be determined. Similarly, the dependence on the polymer flexibility has to be resolved.  }

In conclusion, the present results point toward specific conformational and dynamical properties of semiflexible active linear polymers, with characteristic scaling exponents in an intermediate activity-dependent shear-rate regime. Moreover, their properties approach asymptotically the passive polymer limit, with scaling exponents independent of activity. Deeper insight into this behavior calls for further theoretical studies and experimental investigations to underpin our results.

\begin{acknowledgments}
AL acknowledges useful discussions with G. Gompper in the early stages of this work.
Funding from MIUR Project No. PRIN 2020/PFCXPE is acknowledged.
The work of AL was performed under the auspices of GNFM-INdAM.
\end{acknowledgments}

\section*{Data Availability Statement}

The data that support the findings of this study are available from the corresponding
author upon reasonable request.

\appendix

\section{Brownian Multi-Particle-Collision method}
\label{app:method}

\REV{In the Brownian Multiparticle Collision Dynamics (B-MPC) approach, a bead
stochastically collides with virtual solvent particles in such a way that 
momentum is not conserved, i.e., no hydrodynamic interactions are present~\cite{kiku:03,ripo:07,lamu:24}.
In the B-MPC variant of Ref.~\onlinecite{ripo:07}, the Cartesian components of the linear momentum, ${\bm P}(t)=(P_x(t),P_y(t))^T$, of a single virtual particle of mass $\gamma m$, are sampled from a Maxwell-Boltzmann (MB) distribution, with prescribed mean and variance $\gamma m k_B T$. Shear along the $x$-axis is implemented by the mean value $\langle P_x \rangle = \gamma m \dot{\gamma} y$, and  $\langle P_y \rangle =0$.
Collisions between an ''immersed'' particle---here a bead--- and solvent particles are carried out by the stochastic rotation dynamics variant
of the MPC method~\cite{kapr:08,gomp:09}. The collisions occur at time intervals $\Delta t > \Delta t_p$, where the latter is the time step of the integration scheme for Newton's equations of motion.
In two spatial dimensions, the bead velocity, after the collision with the virtual solvent particle, is given by
\begin{equation} \label{app:mpc_step}
  {\bm v}(t+\Delta t)={\bm v}_{cm}(t+\Delta t)+
  \Omega(\alpha) [{\bm v}'(t+\Delta t)-{\bm v}_{cm}(t+\Delta t)] ,
\end{equation}
where $\Omega(\alpha)$ is the rotation matrix by the angles $\pm \alpha$ with equal probability. ${\bm v}'(t+\Delta t)$ is the velocity of a bead after $\Delta t/\Delta t_p$ simulation steps and ${\bm v}_{cm}(t)$ is the center-of-mass
velocity of the bead and the associated solvent particle,
\begin{equation}
  {\bm v}_{cm}(t)= \frac{M {\bm v}'(t) + {\bm P}(t)}{M+\gamma m} .
\end{equation}  }

\REV{To derive a friction coefficient, we consider the dynamics of an individual force-free bead. As it moves ballistically between collisions, its velocity obeys ${\bm v}'(t+\Delta t) = \bm v(t)$. Averaging of Eq.~\eqref{app:mpc_step} over the stochastic rotation yields in two dimensions:
\begin{equation}
\lla {\bm v}(t+\Delta t) \rra = \lla \bm v(t) \rra +  \frac{\gamma m}{\gamma m +M} (\cos \alpha -1) \lla \bm v(t)  \rra ,
\end{equation}
or 
\begin{equation}
M \frac{\lla {\bm v}(t+\Delta t) - \bm v(t)  \rra}{\Delta t} =  - \xi \lla \bm v(t)  \rra \,
\end{equation}
with the friction coefficient (see also Ref.~\onlinecite{kiku:03}) ($1-\cos \alpha \geq 0$)
\begin{equation} \label{app:friction}
\xi= \frac{m \gamma}{m \gamma+ M} \frac{M}{\Delta t} (1-\cos \alpha) .
\end{equation}
For a proper momentum exchange between beads and solvent particles during a collision, $M \approx \gamma m$.\cite{kiku:03,gomp:09}} 

\section{Supplementary files}
\label{app:movies}

The supplementary multimedia files movie1.mp4, movie2.mp4, and movie3.mp4 available online show animations
of a polymer with $L_p/L=0.4$, $Pe=10^3$ and $Wi=4$, $Wi=70$, and $Wi=10^3$,
respectively. 
The polymer center-of-mass is fixed, the leading bead is depicted in blue, and
the yellow points show a spatially fixed reference
frame as a guide to the eye.


\begin{thebibliography}{93}%
\makeatletter
\providecommand \@ifxundefined [1]{%
 \@ifx{#1\undefined}
}%
\providecommand \@ifnum [1]{%
 \ifnum #1\expandafter \@firstoftwo
 \else \expandafter \@secondoftwo
 \fi
}%
\providecommand \@ifx [1]{%
 \ifx #1\expandafter \@firstoftwo
 \else \expandafter \@secondoftwo
 \fi
}%
\providecommand \natexlab [1]{#1}%
\providecommand \enquote  [1]{``#1''}%
\providecommand \bibnamefont  [1]{#1}%
\providecommand \bibfnamefont [1]{#1}%
\providecommand \citenamefont [1]{#1}%
\providecommand \href@noop [0]{\@secondoftwo}%
\providecommand \href [0]{\begingroup \@sanitize@url \@href}%
\providecommand \@href[1]{\@@startlink{#1}\@@href}%
\providecommand \@@href[1]{\endgroup#1\@@endlink}%
\providecommand \@sanitize@url [0]{\catcode `\\12\catcode `\$12\catcode
  `\&12\catcode `\#12\catcode `\^12\catcode `\_12\catcode `\%12\relax}%
\providecommand \@@startlink[1]{}%
\providecommand \@@endlink[0]{}%
\providecommand \url  [0]{\begingroup\@sanitize@url \@url }%
\providecommand \@url [1]{\endgroup\@href {#1}{\urlprefix }}%
\providecommand \urlprefix  [0]{URL }%
\providecommand \Eprint [0]{\href }%
\providecommand \doibase [0]{https://doi.org/}%
\providecommand \selectlanguage [0]{\@gobble}%
\providecommand \bibinfo  [0]{\@secondoftwo}%
\providecommand \bibfield  [0]{\@secondoftwo}%
\providecommand \translation [1]{[#1]}%
\providecommand \BibitemOpen [0]{}%
\providecommand \bibitemStop [0]{}%
\providecommand \bibitemNoStop [0]{.\EOS\space}%
\providecommand \EOS [0]{\spacefactor3000\relax}%
\providecommand \BibitemShut  [1]{\csname bibitem#1\endcsname}%
\let\auto@bib@innerbib\@empty
\bibitem [{\citenamefont {Winkler}\ and\ \citenamefont
  {Gompper}(2020)}]{wink:20}%
  \BibitemOpen
  \bibfield  {author} {\bibinfo {author} {\bibfnamefont {R.~G.}\ \bibnamefont
  {Winkler}}\ and\ \bibinfo {author} {\bibfnamefont {G.}~\bibnamefont
  {Gompper}},\ }\bibfield  {title} {\enquote {\bibinfo {title} {The physics of
  active polymers and filaments},}\ }\href@noop {} {\bibfield  {journal}
  {\bibinfo  {journal} {J. Chem. Phys.}\ }\textbf {\bibinfo {volume} {153}},\
  \bibinfo {pages} {040901} (\bibinfo {year} {2020})}\BibitemShut {NoStop}%
\bibitem [{\citenamefont {Alberts}\ \emph {et~al.}(2022)\citenamefont
  {Alberts}, \citenamefont {Heald}, \citenamefont {Johnson}, \citenamefont
  {Morgan}, \citenamefont {Raff}, \citenamefont {Roberts},\ and\ \citenamefont
  {Walter}}]{albe:22}%
  \BibitemOpen
  \bibfield  {author} {\bibinfo {author} {\bibfnamefont {B.}~\bibnamefont
  {Alberts}}, \bibinfo {author} {\bibfnamefont {R.}~\bibnamefont {Heald}},
  \bibinfo {author} {\bibfnamefont {A.}~\bibnamefont {Johnson}}, \bibinfo
  {author} {\bibfnamefont {D.}~\bibnamefont {Morgan}}, \bibinfo {author}
  {\bibfnamefont {M.}~\bibnamefont {Raff}}, \bibinfo {author} {\bibfnamefont
  {K.}~\bibnamefont {Roberts}},\ and\ \bibinfo {author} {\bibfnamefont
  {P.}~\bibnamefont {Walter}},\ }\href@noop {} {\emph {\bibinfo {title}
  {Molecular biology of the cell: seventh international student edition with
  registration card}}}\ (\bibinfo  {publisher} {WW Norton \& Company},\
  \bibinfo {year} {2022})\BibitemShut {NoStop}%
\bibitem [{\citenamefont {Guthold}\ \emph {et~al.}(1999)\citenamefont
  {Guthold}, \citenamefont {Zhu}, \citenamefont {Rivetti}, \citenamefont
  {Yang}, \citenamefont {Thomson}, \citenamefont {Kasas}, \citenamefont
  {Hansma}, \citenamefont {Smith}, \citenamefont {Hansma},\ and\ \citenamefont
  {Bustamante}}]{guth:99}%
  \BibitemOpen
  \bibfield  {author} {\bibinfo {author} {\bibfnamefont {M.}~\bibnamefont
  {Guthold}}, \bibinfo {author} {\bibfnamefont {X.}~\bibnamefont {Zhu}},
  \bibinfo {author} {\bibfnamefont {C.}~\bibnamefont {Rivetti}}, \bibinfo
  {author} {\bibfnamefont {G.}~\bibnamefont {Yang}}, \bibinfo {author}
  {\bibfnamefont {N.~H.}\ \bibnamefont {Thomson}}, \bibinfo {author}
  {\bibfnamefont {S.}~\bibnamefont {Kasas}}, \bibinfo {author} {\bibfnamefont
  {H.~G.}\ \bibnamefont {Hansma}}, \bibinfo {author} {\bibfnamefont
  {B.}~\bibnamefont {Smith}}, \bibinfo {author} {\bibfnamefont {P.~K.}\
  \bibnamefont {Hansma}},\ and\ \bibinfo {author} {\bibfnamefont
  {C.}~\bibnamefont {Bustamante}},\ }\bibfield  {title} {\enquote {\bibinfo
  {title} {Direct observation of one-dimensional diffusion and transcription by
  \emph{{E}scherichia coli} {RNA} polymerase},}\ }\href
  {https://doi.org/https://doi.org/10.1016/S0006-3495(99)77067-0} {\bibfield
  {journal} {\bibinfo  {journal} {Biophys. J.}\ }\textbf {\bibinfo {volume}
  {77}},\ \bibinfo {pages} {2284} (\bibinfo {year} {1999})}\BibitemShut
  {NoStop}%
\bibitem [{\citenamefont {Mejia}, \citenamefont {Nudler},\ and\ \citenamefont
  {Bustamante}(2015)}]{meji:15}%
  \BibitemOpen
  \bibfield  {author} {\bibinfo {author} {\bibfnamefont {Y.~X.}\ \bibnamefont
  {Mejia}}, \bibinfo {author} {\bibfnamefont {E.}~\bibnamefont {Nudler}},\ and\
  \bibinfo {author} {\bibfnamefont {C.}~\bibnamefont {Bustamante}},\ }\bibfield
   {title} {\enquote {\bibinfo {title} {Trigger loop folding determines
  transcription rate of \emph{{E}scherichia coli's} {RNA} polymerase},}\ }\href
  {https://doi.org/10.1073/pnas.1421067112} {\bibfield  {journal} {\bibinfo
  {journal} {Proc. Natl. Acad. Sci. USA}\ }\textbf {\bibinfo {volume} {112}},\
  \bibinfo {pages} {743} (\bibinfo {year} {2015})}\BibitemShut {NoStop}%
\bibitem [{\citenamefont {Di~Pierro}\ \emph {et~al.}(2018)\citenamefont
  {Di~Pierro}, \citenamefont {Potoyan}, \citenamefont {Wolynes},\ and\
  \citenamefont {Onuchic}}]{dipi:18}%
  \BibitemOpen
  \bibfield  {author} {\bibinfo {author} {\bibfnamefont {M.}~\bibnamefont
  {Di~Pierro}}, \bibinfo {author} {\bibfnamefont {D.~A.}\ \bibnamefont
  {Potoyan}}, \bibinfo {author} {\bibfnamefont {P.~G.}\ \bibnamefont
  {Wolynes}},\ and\ \bibinfo {author} {\bibfnamefont {J.~N.}\ \bibnamefont
  {Onuchic}},\ }\bibfield  {title} {\enquote {\bibinfo {title} {Anomalous
  diffusion, spatial coherence, and viscoelasticity from the energy landscape
  of human chromosomes},}\ }\href {https://doi.org/10.1073/pnas.1806297115}
  {\bibfield  {journal} {\bibinfo  {journal} {Proc. Natl. Acad. Sci. USA}\
  }\textbf {\bibinfo {volume} {115}},\ \bibinfo {pages} {7753} (\bibinfo {year}
  {2018})}\BibitemShut {NoStop}%
\bibitem [{\citenamefont {Javer}\ \emph {et~al.}(2013)\citenamefont {Javer},
  \citenamefont {Long}, \citenamefont {Nugent}, \citenamefont {Grisi},
  \citenamefont {Siriwatwetchakul}, \citenamefont {Dorfman}, \citenamefont
  {Cicuta},\ and\ \citenamefont {Cosentino~Lagomarsino}}]{jave:13}%
  \BibitemOpen
  \bibfield  {author} {\bibinfo {author} {\bibfnamefont {A.}~\bibnamefont
  {Javer}}, \bibinfo {author} {\bibfnamefont {Z.}~\bibnamefont {Long}},
  \bibinfo {author} {\bibfnamefont {E.}~\bibnamefont {Nugent}}, \bibinfo
  {author} {\bibfnamefont {M.}~\bibnamefont {Grisi}}, \bibinfo {author}
  {\bibfnamefont {K.}~\bibnamefont {Siriwatwetchakul}}, \bibinfo {author}
  {\bibfnamefont {K.~D.}\ \bibnamefont {Dorfman}}, \bibinfo {author}
  {\bibfnamefont {P.}~\bibnamefont {Cicuta}},\ and\ \bibinfo {author}
  {\bibfnamefont {M.}~\bibnamefont {Cosentino~Lagomarsino}},\ }\bibfield
  {title} {\enquote {\bibinfo {title} {Short-time movement of {{\em {E} coli}}
  chromosomal loci depends on coordinate and subcellular localization},}\
  }\href {https://doi.org/10.1038/ncomms3003} {\bibfield  {journal} {\bibinfo
  {journal} {Nat. Commun.}\ }\textbf {\bibinfo {volume} {4}},\ \bibinfo {pages}
  {3003} (\bibinfo {year} {2013})}\BibitemShut {NoStop}%
\bibitem [{\citenamefont {Zidovska}, \citenamefont {Weitz},\ and\ \citenamefont
  {Mitchison}(2013)}]{zido:13}%
  \BibitemOpen
  \bibfield  {author} {\bibinfo {author} {\bibfnamefont {A.}~\bibnamefont
  {Zidovska}}, \bibinfo {author} {\bibfnamefont {D.~A.}\ \bibnamefont
  {Weitz}},\ and\ \bibinfo {author} {\bibfnamefont {T.~J.}\ \bibnamefont
  {Mitchison}},\ }\bibfield  {title} {\enquote {\bibinfo {title} {Micron-scale
  coherence in interphase chromatin dynamics},}\ }\href
  {http://www.pnas.org/content/110/39/15555.abstract} {\bibfield  {journal}
  {\bibinfo  {journal} {Proc. Natl. Acad. Sci. USA}\ }\textbf {\bibinfo
  {volume} {110}},\ \bibinfo {pages} {15555} (\bibinfo {year}
  {2013})}\BibitemShut {NoStop}%
\bibitem [{\citenamefont {Lieberman-Aiden}\ \emph {et~al.}(2009)\citenamefont
  {Lieberman-Aiden}, \citenamefont {van Berkum}, \citenamefont {Williams},
  \citenamefont {Imakaev}, \citenamefont {Ragoczy}, \citenamefont {Telling},
  \citenamefont {Amit}, \citenamefont {Lajoie}, \citenamefont {Sabo},
  \citenamefont {Dorschner}, \citenamefont {Sandstrom}, \citenamefont
  {Bernstein}, \citenamefont {Bender}, \citenamefont {Groudine}, \citenamefont
  {Gnirke}, \citenamefont {Stamatoyannopoulos}, \citenamefont {Mirny},
  \citenamefont {Lander},\ and\ \citenamefont {Dekker}}]{lieb:09}%
  \BibitemOpen
  \bibfield  {author} {\bibinfo {author} {\bibfnamefont {E.}~\bibnamefont
  {Lieberman-Aiden}}, \bibinfo {author} {\bibfnamefont {N.~L.}\ \bibnamefont
  {van Berkum}}, \bibinfo {author} {\bibfnamefont {L.}~\bibnamefont
  {Williams}}, \bibinfo {author} {\bibfnamefont {M.}~\bibnamefont {Imakaev}},
  \bibinfo {author} {\bibfnamefont {T.}~\bibnamefont {Ragoczy}}, \bibinfo
  {author} {\bibfnamefont {A.}~\bibnamefont {Telling}}, \bibinfo {author}
  {\bibfnamefont {I.}~\bibnamefont {Amit}}, \bibinfo {author} {\bibfnamefont
  {B.~R.}\ \bibnamefont {Lajoie}}, \bibinfo {author} {\bibfnamefont {P.~J.}\
  \bibnamefont {Sabo}}, \bibinfo {author} {\bibfnamefont {M.~O.}\ \bibnamefont
  {Dorschner}}, \bibinfo {author} {\bibfnamefont {R.}~\bibnamefont
  {Sandstrom}}, \bibinfo {author} {\bibfnamefont {B.}~\bibnamefont
  {Bernstein}}, \bibinfo {author} {\bibfnamefont {M.~A.}\ \bibnamefont
  {Bender}}, \bibinfo {author} {\bibfnamefont {M.}~\bibnamefont {Groudine}},
  \bibinfo {author} {\bibfnamefont {A.}~\bibnamefont {Gnirke}}, \bibinfo
  {author} {\bibfnamefont {J.}~\bibnamefont {Stamatoyannopoulos}}, \bibinfo
  {author} {\bibfnamefont {L.~A.}\ \bibnamefont {Mirny}}, \bibinfo {author}
  {\bibfnamefont {E.~S.}\ \bibnamefont {Lander}},\ and\ \bibinfo {author}
  {\bibfnamefont {J.}~\bibnamefont {Dekker}},\ }\bibfield  {title} {\enquote
  {\bibinfo {title} {Comprehensive mapping of long-range interactions reveals
  folding principles of the human genome},}\ }\href
  {https://doi.org/10.1126/science.1181369} {\bibfield  {journal} {\bibinfo
  {journal} {Science}\ }\textbf {\bibinfo {volume} {326}},\ \bibinfo {pages}
  {289} (\bibinfo {year} {2009})}\BibitemShut {NoStop}%
\bibitem [{\citenamefont {Ganai}, \citenamefont {Sengupta},\ and\ \citenamefont
  {Menon}(2014)}]{gana:14}%
  \BibitemOpen
  \bibfield  {author} {\bibinfo {author} {\bibfnamefont {N.}~\bibnamefont
  {Ganai}}, \bibinfo {author} {\bibfnamefont {S.}~\bibnamefont {Sengupta}},\
  and\ \bibinfo {author} {\bibfnamefont {G.~I.}\ \bibnamefont {Menon}},\
  }\bibfield  {title} {\enquote {\bibinfo {title} {Chromosome positioning from
  activity-based segregation},}\ }\href {https://doi.org/10.1093/nar/gkt1417}
  {\bibfield  {journal} {\bibinfo  {journal} {Nucleic Acids Res.}\ }\textbf
  {\bibinfo {volume} {42}},\ \bibinfo {pages} {4145} (\bibinfo {year}
  {2014})}\BibitemShut {NoStop}%
\bibitem [{\citenamefont {Saintillan}, \citenamefont {Shelley},\ and\
  \citenamefont {Zidovska}(2018)}]{sain:18.1}%
  \BibitemOpen
  \bibfield  {author} {\bibinfo {author} {\bibfnamefont {D.}~\bibnamefont
  {Saintillan}}, \bibinfo {author} {\bibfnamefont {M.~J.}\ \bibnamefont
  {Shelley}},\ and\ \bibinfo {author} {\bibfnamefont {A.}~\bibnamefont
  {Zidovska}},\ }\bibfield  {title} {\enquote {\bibinfo {title} {Extensile
  motor activity drives coherent motions in a model of interphase chromatin},}\
  }\href {https://doi.org/10.1073/pnas.1807073115} {\bibfield  {journal}
  {\bibinfo  {journal} {Proc. Natl. Acad. Sci. USA}\ }\textbf {\bibinfo
  {volume} {115}},\ \bibinfo {pages} {11442} (\bibinfo {year}
  {2018})}\BibitemShut {NoStop}%
\bibitem [{\citenamefont {Goychuk}\ \emph {et~al.}(2023)\citenamefont
  {Goychuk}, \citenamefont {Kannan}, \citenamefont {Chakraborty},\ and\
  \citenamefont {Kardar}}]{goyc:23}%
  \BibitemOpen
  \bibfield  {author} {\bibinfo {author} {\bibfnamefont {A.}~\bibnamefont
  {Goychuk}}, \bibinfo {author} {\bibfnamefont {D.}~\bibnamefont {Kannan}},
  \bibinfo {author} {\bibfnamefont {A.~K.}\ \bibnamefont {Chakraborty}},\ and\
  \bibinfo {author} {\bibfnamefont {M.}~\bibnamefont {Kardar}},\ }\bibfield
  {title} {\enquote {\bibinfo {title} {Polymer folding through active processes
  recreates features of genome organization},}\ }\href
  {https://doi.org/10.1073/pnas.2221726120} {\bibfield  {journal} {\bibinfo
  {journal} {Proc. Natl. Acad. Sci. USA}\ }\textbf {\bibinfo {volume} {120}},\
  \bibinfo {pages} {e2221726120} (\bibinfo {year} {2023})}\BibitemShut
  {NoStop}%
\bibitem [{\citenamefont {Smrek}\ and\ \citenamefont {Kremer}(2017)}]{smre:17}%
  \BibitemOpen
  \bibfield  {author} {\bibinfo {author} {\bibfnamefont {J.}~\bibnamefont
  {Smrek}}\ and\ \bibinfo {author} {\bibfnamefont {K.}~\bibnamefont {Kremer}},\
  }\bibfield  {title} {\enquote {\bibinfo {title} {Small activity differences
  drive phase separation in active-passive polymer mixtures},}\ }\href
  {https://doi.org/10.1103/PhysRevLett.118.098002} {\bibfield  {journal}
  {\bibinfo  {journal} {Phys. Rev. Lett.}\ }\textbf {\bibinfo {volume} {118}},\
  \bibinfo {pages} {098002} (\bibinfo {year} {2017})}\BibitemShut {NoStop}%
\bibitem [{\citenamefont {N{\'e}d{\'e}lec}\ \emph {et~al.}(1997)\citenamefont
  {N{\'e}d{\'e}lec}, \citenamefont {Surrey}, \citenamefont {Maggs},\ and\
  \citenamefont {Leibler}}]{nedl:97}%
  \BibitemOpen
  \bibfield  {author} {\bibinfo {author} {\bibfnamefont {F.~J.}\ \bibnamefont
  {N{\'e}d{\'e}lec}}, \bibinfo {author} {\bibfnamefont {T.}~\bibnamefont
  {Surrey}}, \bibinfo {author} {\bibfnamefont {A.~C.}\ \bibnamefont {Maggs}},\
  and\ \bibinfo {author} {\bibfnamefont {S.}~\bibnamefont {Leibler}},\
  }\bibfield  {title} {\enquote {\bibinfo {title} {Self-organization of
  microtubules and motors},}\ }\href {https://doi.org/10.1038/38532} {\bibfield
   {journal} {\bibinfo  {journal} {Nature}\ }\textbf {\bibinfo {volume}
  {389}},\ \bibinfo {pages} {305} (\bibinfo {year} {1997})}\BibitemShut
  {NoStop}%
\bibitem [{\citenamefont {Brangwynne}\ \emph {et~al.}(2008)\citenamefont
  {Brangwynne}, \citenamefont {Koenderink}, \citenamefont {MacKintosh},\ and\
  \citenamefont {Weitz}}]{bran:08}%
  \BibitemOpen
  \bibfield  {author} {\bibinfo {author} {\bibfnamefont {C.~P.}\ \bibnamefont
  {Brangwynne}}, \bibinfo {author} {\bibfnamefont {G.~H.}\ \bibnamefont
  {Koenderink}}, \bibinfo {author} {\bibfnamefont {F.~C.}\ \bibnamefont
  {MacKintosh}},\ and\ \bibinfo {author} {\bibfnamefont {D.~A.}\ \bibnamefont
  {Weitz}},\ }\bibfield  {title} {\enquote {\bibinfo {title} {Cytoplasmic
  diffusion: molecular motors mix it up},}\ }\href
  {https://doi.org/10.1083/jcb.200806149} {\bibfield  {journal} {\bibinfo
  {journal} {J. Cell. Biol.}\ }\textbf {\bibinfo {volume} {183}},\ \bibinfo
  {pages} {583} (\bibinfo {year} {2008})}\BibitemShut {NoStop}%
\bibitem [{\citenamefont {Weber}\ \emph {et~al.}(2015)\citenamefont {Weber},
  \citenamefont {Suzuki}, \citenamefont {Schaller}, \citenamefont {Aranson},
  \citenamefont {Bausch},\ and\ \citenamefont {Frey}}]{webe:15}%
  \BibitemOpen
  \bibfield  {author} {\bibinfo {author} {\bibfnamefont {C.~A.}\ \bibnamefont
  {Weber}}, \bibinfo {author} {\bibfnamefont {R.}~\bibnamefont {Suzuki}},
  \bibinfo {author} {\bibfnamefont {V.}~\bibnamefont {Schaller}}, \bibinfo
  {author} {\bibfnamefont {I.~S.}\ \bibnamefont {Aranson}}, \bibinfo {author}
  {\bibfnamefont {A.~R.}\ \bibnamefont {Bausch}},\ and\ \bibinfo {author}
  {\bibfnamefont {E.}~\bibnamefont {Frey}},\ }\bibfield  {title} {\enquote
  {\bibinfo {title} {Random bursts determine dynamics of active filaments},}\
  }\href {https://doi.org/10.1073/pnas.1421322112} {\bibfield  {journal}
  {\bibinfo  {journal} {Proc. Natl. Acad. Sci. USA}\ }\textbf {\bibinfo
  {volume} {112}},\ \bibinfo {pages} {10703} (\bibinfo {year}
  {2015})}\BibitemShut {NoStop}%
\bibitem [{\citenamefont {Lau}\ \emph {et~al.}(2003)\citenamefont {Lau},
  \citenamefont {Hoffman}, \citenamefont {Davies}, \citenamefont {Crocker},\
  and\ \citenamefont {Lubensky}}]{lau:03}%
  \BibitemOpen
  \bibfield  {author} {\bibinfo {author} {\bibfnamefont {A.~W.~C.}\
  \bibnamefont {Lau}}, \bibinfo {author} {\bibfnamefont {B.~D.}\ \bibnamefont
  {Hoffman}}, \bibinfo {author} {\bibfnamefont {A.}~\bibnamefont {Davies}},
  \bibinfo {author} {\bibfnamefont {J.~C.}\ \bibnamefont {Crocker}},\ and\
  \bibinfo {author} {\bibfnamefont {T.~C.}\ \bibnamefont {Lubensky}},\
  }\bibfield  {title} {\enquote {\bibinfo {title} {Microrheology, stress
  fluctuations, and active behavior of living cells},}\ }\href
  {https://doi.org/10.1103/PhysRevLett.91.198101} {\bibfield  {journal}
  {\bibinfo  {journal} {Phys. Rev. Lett.}\ }\textbf {\bibinfo {volume} {91}},\
  \bibinfo {pages} {198101} (\bibinfo {year} {2003})}\BibitemShut {NoStop}%
\bibitem [{\citenamefont {MacKintosh}\ and\ \citenamefont
  {Levine}(2008)}]{mack:08}%
  \BibitemOpen
  \bibfield  {author} {\bibinfo {author} {\bibfnamefont {F.~C.}\ \bibnamefont
  {MacKintosh}}\ and\ \bibinfo {author} {\bibfnamefont {A.~J.}\ \bibnamefont
  {Levine}},\ }\bibfield  {title} {\enquote {\bibinfo {title} {Nonequilibrium
  mechanics and dynamics of motor-activated gels},}\ }\href
  {https://doi.org/10.1103/PhysRevLett.100.018104} {\bibfield  {journal}
  {\bibinfo  {journal} {Phys. Rev. Lett.}\ }\textbf {\bibinfo {volume} {100}},\
  \bibinfo {pages} {018104} (\bibinfo {year} {2008})}\BibitemShut {NoStop}%
\bibitem [{\citenamefont {Lu}\ \emph {et~al.}(2016)\citenamefont {Lu},
  \citenamefont {Winding}, \citenamefont {Lakonishok}, \citenamefont
  {Wildonger},\ and\ \citenamefont {Gelfand}}]{lu:16}%
  \BibitemOpen
  \bibfield  {author} {\bibinfo {author} {\bibfnamefont {W.}~\bibnamefont
  {Lu}}, \bibinfo {author} {\bibfnamefont {M.}~\bibnamefont {Winding}},
  \bibinfo {author} {\bibfnamefont {M.}~\bibnamefont {Lakonishok}}, \bibinfo
  {author} {\bibfnamefont {J.}~\bibnamefont {Wildonger}},\ and\ \bibinfo
  {author} {\bibfnamefont {V.~I.}\ \bibnamefont {Gelfand}},\ }\bibfield
  {title} {\enquote {\bibinfo {title} {Microtubule--microtubule sliding by
  kinesin-1 is essential for normal cytoplasmic streaming in
  \emph{{D}rosophila} oocytes},}\ }\href
  {https://doi.org/10.1073/pnas.1522424113} {\bibfield  {journal} {\bibinfo
  {journal} {Proc. Natl. Acad. Sci. USA}\ }\textbf {\bibinfo {volume} {113}},\
  \bibinfo {pages} {E4995} (\bibinfo {year} {2016})}\BibitemShut {NoStop}%
\bibitem [{\citenamefont {Ravichandran}\ \emph {et~al.}(2017)\citenamefont
  {Ravichandran}, \citenamefont {Vliegenthart}, \citenamefont {Saggiorato},
  \citenamefont {Auth},\ and\ \citenamefont {Gompper}}]{ravi:17}%
  \BibitemOpen
  \bibfield  {author} {\bibinfo {author} {\bibfnamefont {A.}~\bibnamefont
  {Ravichandran}}, \bibinfo {author} {\bibfnamefont {G.~A.}\ \bibnamefont
  {Vliegenthart}}, \bibinfo {author} {\bibfnamefont {G.}~\bibnamefont
  {Saggiorato}}, \bibinfo {author} {\bibfnamefont {T.}~\bibnamefont {Auth}},\
  and\ \bibinfo {author} {\bibfnamefont {G.}~\bibnamefont {Gompper}},\
  }\bibfield  {title} {\enquote {\bibinfo {title} {Enhanced dynamics of
  confined cytoskeletal filaments driven by asymmetric motors},}\ }\href
  {https://doi.org/https://doi.org/10.1016/j.bpj.2017.07.016} {\bibfield
  {journal} {\bibinfo  {journal} {Biophys. J.}\ }\textbf {\bibinfo {volume}
  {113}},\ \bibinfo {pages} {1121} (\bibinfo {year} {2017})}\BibitemShut
  {NoStop}%
\bibitem [{\citenamefont {Nguyen}\ \emph {et~al.}(2021)\citenamefont {Nguyen},
  \citenamefont {Ozkan-Aydin}, \citenamefont {Tuazon}, \citenamefont {Goldman},
  \citenamefont {Bhamla},\ and\ \citenamefont {Peleg}}]{nguy:21}%
  \BibitemOpen
  \bibfield  {author} {\bibinfo {author} {\bibfnamefont {C.}~\bibnamefont
  {Nguyen}}, \bibinfo {author} {\bibfnamefont {Y.}~\bibnamefont {Ozkan-Aydin}},
  \bibinfo {author} {\bibfnamefont {H.}~\bibnamefont {Tuazon}}, \bibinfo
  {author} {\bibfnamefont {D.~I.}\ \bibnamefont {Goldman}}, \bibinfo {author}
  {\bibfnamefont {M.~S.}\ \bibnamefont {Bhamla}},\ and\ \bibinfo {author}
  {\bibfnamefont {O.}~\bibnamefont {Peleg}},\ }\bibfield  {title} {\enquote
  {\bibinfo {title} {Emergent collective locomotion in an active polymer model
  of entangled worm blobs},}\ }\href {https://doi.org/10.3389/fphy.2021.734499}
  {\bibfield  {journal} {\bibinfo  {journal} {Front. Phys.}\ }\textbf {\bibinfo
  {volume} {9}},\ \bibinfo {pages} {734499} (\bibinfo {year}
  {2021})}\BibitemShut {NoStop}%
\bibitem [{\citenamefont {Gagnon}\ and\ \citenamefont
  {Arratia}(2016)}]{gagn:16}%
  \BibitemOpen
  \bibfield  {author} {\bibinfo {author} {\bibfnamefont {D.~A.}\ \bibnamefont
  {Gagnon}}\ and\ \bibinfo {author} {\bibfnamefont {P.~E.}\ \bibnamefont
  {Arratia}},\ }\bibfield  {title} {\enquote {\bibinfo {title} {The cost of
  swimming in generalized newtonian fluids: experiments with \emph{{C}.
  elegans}},}\ }\href {https://doi.org/DOI: 10.1017/jfm.2016.420} {\bibfield
  {journal} {\bibinfo  {journal} {J. Fluid Mech.}\ }\textbf {\bibinfo {volume}
  {800}},\ \bibinfo {pages} {753} (\bibinfo {year} {2016})}\BibitemShut
  {NoStop}%
\bibitem [{\citenamefont {Ding}\ \emph {et~al.}(2019)\citenamefont {Ding},
  \citenamefont {Schumacher}, \citenamefont {Javer}, \citenamefont {Endres},
  \citenamefont {Brown}, \citenamefont {Barkai}, \citenamefont {Mignot},
  \citenamefont {Hodgkin},\ and\ \citenamefont {Igoshin}}]{ding:19}%
  \BibitemOpen
  \bibfield  {author} {\bibinfo {author} {\bibfnamefont {S.~S.}\ \bibnamefont
  {Ding}}, \bibinfo {author} {\bibfnamefont {L.~J.}\ \bibnamefont
  {Schumacher}}, \bibinfo {author} {\bibfnamefont {A.~E.}\ \bibnamefont
  {Javer}}, \bibinfo {author} {\bibfnamefont {R.~G.}\ \bibnamefont {Endres}},
  \bibinfo {author} {\bibfnamefont {A.}~\bibnamefont {Brown}}, \bibinfo
  {author} {\bibfnamefont {N.}~\bibnamefont {Barkai}}, \bibinfo {author}
  {\bibfnamefont {T.}~\bibnamefont {Mignot}}, \bibinfo {author} {\bibfnamefont
  {J.}~\bibnamefont {Hodgkin}},\ and\ \bibinfo {author} {\bibfnamefont {O.~A.}\
  \bibnamefont {Igoshin}},\ }\bibfield  {title} {\enquote {\bibinfo {title}
  {Shared behavioral mechanisms underlie c. elegans aggregation and
  swarming},}\ }\href {https://doi.org/10.7554/eLife.43318} {\bibfield
  {journal} {\bibinfo  {journal} {eLife}\ }\textbf {\bibinfo {volume} {8}},\
  \bibinfo {pages} {e43318} (\bibinfo {year} {2019})}\BibitemShut {NoStop}%
\bibitem [{\citenamefont {Deblais}, \citenamefont {Woutersen},\ and\
  \citenamefont {Bonn}(2020)}]{debl:20}%
  \BibitemOpen
  \bibfield  {author} {\bibinfo {author} {\bibfnamefont {A.}~\bibnamefont
  {Deblais}}, \bibinfo {author} {\bibfnamefont {S.}~\bibnamefont {Woutersen}},\
  and\ \bibinfo {author} {\bibfnamefont {D.}~\bibnamefont {Bonn}},\ }\bibfield
  {title} {\enquote {\bibinfo {title} {Rheology of entangled active
  polymer-like {T. Tubifex} worms},}\ }\href
  {https://doi.org/10.1103/PhysRevLett.124.188002} {\bibfield  {journal}
  {\bibinfo  {journal} {Phys. Rev. Lett.}\ }\textbf {\bibinfo {volume} {124}},\
  \bibinfo {pages} {188002} (\bibinfo {year} {2020})}\BibitemShut {NoStop}%
\bibitem [{\citenamefont {Deblais}\ \emph {et~al.}(2020)\citenamefont
  {Deblais}, \citenamefont {Maggs}, \citenamefont {Bonn},\ and\ \citenamefont
  {Woutersen}}]{debl:20.1}%
  \BibitemOpen
  \bibfield  {author} {\bibinfo {author} {\bibfnamefont {A.}~\bibnamefont
  {Deblais}}, \bibinfo {author} {\bibfnamefont {A.~C.}\ \bibnamefont {Maggs}},
  \bibinfo {author} {\bibfnamefont {D.}~\bibnamefont {Bonn}},\ and\ \bibinfo
  {author} {\bibfnamefont {S.}~\bibnamefont {Woutersen}},\ }\bibfield  {title}
  {\enquote {\bibinfo {title} {Phase separation by entanglement of active
  polymerlike worms},}\ }\href {https://doi.org/10.1103/PhysRevLett.124.208006}
  {\bibfield  {journal} {\bibinfo  {journal} {Phys. Rev. Lett.}\ }\textbf
  {\bibinfo {volume} {124}},\ \bibinfo {pages} {208006} (\bibinfo {year}
  {2020})}\BibitemShut {NoStop}%
\bibitem [{\citenamefont {Sugi}\ \emph {et~al.}(2019)\citenamefont {Sugi},
  \citenamefont {Ito}, \citenamefont {Nishimura},\ and\ \citenamefont
  {Nagai}}]{sugi:19}%
  \BibitemOpen
  \bibfield  {author} {\bibinfo {author} {\bibfnamefont {T.}~\bibnamefont
  {Sugi}}, \bibinfo {author} {\bibfnamefont {H.}~\bibnamefont {Ito}}, \bibinfo
  {author} {\bibfnamefont {M.}~\bibnamefont {Nishimura}},\ and\ \bibinfo
  {author} {\bibfnamefont {K.~H.}\ \bibnamefont {Nagai}},\ }\bibfield  {title}
  {\enquote {\bibinfo {title} {\emph{{C}. elegans} collectively forms dynamical
  networks},}\ }\href {https://doi.org/10.1038/s41467-019-08537-y} {\bibfield
  {journal} {\bibinfo  {journal} {Nat. Commun.}\ }\textbf {\bibinfo {volume}
  {10}},\ \bibinfo {pages} {683} (\bibinfo {year} {2019})}\BibitemShut
  {NoStop}%
\bibitem [{\citenamefont {Panda}, \citenamefont {Winkler},\ and\ \citenamefont
  {Singh}(2025)}]{pand:25}%
  \BibitemOpen
  \bibfield  {author} {\bibinfo {author} {\bibfnamefont {A.}~\bibnamefont
  {Panda}}, \bibinfo {author} {\bibfnamefont {R.~G.}\ \bibnamefont {Winkler}},\
  and\ \bibinfo {author} {\bibfnamefont {S.~P.}\ \bibnamefont {Singh}},\
  }\bibfield  {title} {\enquote {\bibinfo {title} {Activity-enhanced shear
  thinning of flexible linear polar polymers},}\ }\href@noop {} {\bibfield
  {journal} {\bibinfo  {journal} {Phys. Rev. E}\ }\textbf {\bibinfo {volume}
  {111}},\ \bibinfo {pages} {055413} (\bibinfo {year} {2025})}\BibitemShut
  {NoStop}%
\bibitem [{\citenamefont {Panda}, \citenamefont {Singh},\ and\ \citenamefont
  {Winkler}(2025)}]{pand:25.2}%
  \BibitemOpen
  \bibfield  {author} {\bibinfo {author} {\bibfnamefont {A.}~\bibnamefont
  {Panda}}, \bibinfo {author} {\bibfnamefont {S.~P.}\ \bibnamefont {Singh}},\
  and\ \bibinfo {author} {\bibfnamefont {R.~G.}\ \bibnamefont {Winkler}},\
  }\bibfield  {title} {\enquote {\bibinfo {title} {Analytical analysis of the
  conformational and rheological properties of flexible active polar linear
  polymers under shear flow},}\ }\href {https://doi.org/10.1063/5.0307774}
  {\bibfield  {journal} {\bibinfo  {journal} {J. Chem. Phys.}\ }\textbf
  {\bibinfo {volume} {163}},\ \bibinfo {pages} {224905} (\bibinfo {year}
  {2025})}\BibitemShut {NoStop}%
\bibitem [{\citenamefont {Malvar}, \citenamefont {Carmo},\ and\ \citenamefont
  {Cunha}(2019)}]{malv:19}%
  \BibitemOpen
  \bibfield  {author} {\bibinfo {author} {\bibfnamefont {S.}~\bibnamefont
  {Malvar}}, \bibinfo {author} {\bibfnamefont {B.~S.}\ \bibnamefont {Carmo}},\
  and\ \bibinfo {author} {\bibfnamefont {F.~R.}\ \bibnamefont {Cunha}},\
  }\bibfield  {title} {\enquote {\bibinfo {title} {Rheology of a nematic active
  suspension undergoing oscillatory shear and step strain flows},}\ }\href
  {https://doi.org/10.1007/s00397-019-01178-4} {\bibfield  {journal} {\bibinfo
  {journal} {Rheol. Acta}\ }\textbf {\bibinfo {volume} {58}},\ \bibinfo {pages}
  {771--779} (\bibinfo {year} {2019})}\BibitemShut {NoStop}%
\bibitem [{\citenamefont {Larson}(1999)}]{lars:99}%
  \BibitemOpen
  \bibfield  {author} {\bibinfo {author} {\bibfnamefont {R.~G.}\ \bibnamefont
  {Larson}},\ }\href@noop {} {\emph {\bibinfo {title} {The Structure and
  Rheology of Complex Fluids}}}\ (\bibinfo  {publisher} {Oxford University
  Press},\ \bibinfo {address} {New York},\ \bibinfo {year} {1999})\BibitemShut
  {NoStop}%
\bibitem [{\citenamefont {Xu}\ \emph {et~al.}(2022)\citenamefont {Xu},
  \citenamefont {Sun}, \citenamefont {Lu}, \citenamefont {Patil}, \citenamefont
  {Mays}, \citenamefont {Schweizer},\ and\ \citenamefont {Cheng}}]{xu:22}%
  \BibitemOpen
  \bibfield  {author} {\bibinfo {author} {\bibfnamefont {Z.}~\bibnamefont
  {Xu}}, \bibinfo {author} {\bibfnamefont {R.}~\bibnamefont {Sun}}, \bibinfo
  {author} {\bibfnamefont {W.}~\bibnamefont {Lu}}, \bibinfo {author}
  {\bibfnamefont {S.}~\bibnamefont {Patil}}, \bibinfo {author} {\bibfnamefont
  {J.}~\bibnamefont {Mays}}, \bibinfo {author} {\bibfnamefont {K.~S.}\
  \bibnamefont {Schweizer}},\ and\ \bibinfo {author} {\bibfnamefont
  {S.}~\bibnamefont {Cheng}},\ }\bibfield  {title} {\enquote {\bibinfo {title}
  {Nature of steady-state fast flow in entangled polymer melts: chain
  stretching, shear thinning, and viscosity scaling},}\ }\href@noop {}
  {\bibfield  {journal} {\bibinfo  {journal} {Macromolecules}\ }\textbf
  {\bibinfo {volume} {55}},\ \bibinfo {pages} {10737--10750} (\bibinfo {year}
  {2022})}\BibitemShut {NoStop}%
\bibitem [{\citenamefont {Mart{{\'\i}}n-G{{\'o}}mez}, \citenamefont {Gompper},\
  and\ \citenamefont {Winkler}(2018)}]{mart:18.1}%
  \BibitemOpen
  \bibfield  {author} {\bibinfo {author} {\bibfnamefont {A.}~\bibnamefont
  {Mart{{\'\i}}n-G{{\'o}}mez}}, \bibinfo {author} {\bibfnamefont
  {G.}~\bibnamefont {Gompper}},\ and\ \bibinfo {author} {\bibfnamefont {R.~G.}\
  \bibnamefont {Winkler}},\ }\bibfield  {title} {\enquote {\bibinfo {title}
  {Active {B}rownian filamentous polymers under shear flow},}\ }\href
  {https://doi.org/10.3390/polym10080837} {\bibfield  {journal} {\bibinfo
  {journal} {Polymers}\ }\textbf {\bibinfo {volume} {10}},\ \bibinfo {pages}
  {837} (\bibinfo {year} {2018})}\BibitemShut {NoStop}%
\bibitem [{\citenamefont {Winkler}\ and\ \citenamefont
  {Singh}(2024)}]{wink:24}%
  \BibitemOpen
  \bibfield  {author} {\bibinfo {author} {\bibfnamefont {R.~G.}\ \bibnamefont
  {Winkler}}\ and\ \bibinfo {author} {\bibfnamefont {S.~P.}\ \bibnamefont
  {Singh}},\ }\bibfield  {title} {\enquote {\bibinfo {title} {Active polar ring
  polymer in shear flow - {A}n analytical study},}\ }\href@noop {} {\bibfield
  {journal} {\bibinfo  {journal} {J. Chem. Phys.}\ }\textbf {\bibinfo {volume}
  {161}},\ \bibinfo {pages} {064902} (\bibinfo {year} {2024})}\BibitemShut
  {NoStop}%
\bibitem [{\citenamefont {Panda}, \citenamefont {Winkler},\ and\ \citenamefont
  {Singh}(2023)}]{pand:23}%
  \BibitemOpen
  \bibfield  {author} {\bibinfo {author} {\bibfnamefont {A.}~\bibnamefont
  {Panda}}, \bibinfo {author} {\bibfnamefont {R.~G.}\ \bibnamefont {Winkler}},\
  and\ \bibinfo {author} {\bibfnamefont {S.~P.}\ \bibnamefont {Singh}},\
  }\bibfield  {title} {\enquote {\bibinfo {title} {Characteristic features of
  self-avoiding active {B}rownian polymers under linear shear flow},}\
  }\href@noop {} {\bibfield  {journal} {\bibinfo  {journal} {Soft Matter}\
  }\textbf {\bibinfo {volume} {19}},\ \bibinfo {pages} {8577} (\bibinfo {year}
  {2023})}\BibitemShut {NoStop}%
\bibitem [{\citenamefont {Kumar}, \citenamefont {Padinhateeri},\ and\
  \citenamefont {Thakur}(2024)}]{kuma:24}%
  \BibitemOpen
  \bibfield  {author} {\bibinfo {author} {\bibfnamefont {S.}~\bibnamefont
  {Kumar}}, \bibinfo {author} {\bibfnamefont {R.}~\bibnamefont
  {Padinhateeri}},\ and\ \bibinfo {author} {\bibfnamefont {S.}~\bibnamefont
  {Thakur}},\ }\bibfield  {title} {\enquote {\bibinfo {title} {Shear flow as a
  tool to distinguish microscopic activities of molecular machines in a
  chromatin loop},}\ }\href {https://doi.org/10.1039/D4SM00636D} {\bibfield
  {journal} {\bibinfo  {journal} {Soft Matter}\ }\textbf {\bibinfo {volume}
  {20}},\ \bibinfo {pages} {6500} (\bibinfo {year} {2024})}\BibitemShut
  {NoStop}%
\bibitem [{\citenamefont {Kaiser}\ and\ \citenamefont
  {L{\"o}wen}(2014)}]{kais:14}%
  \BibitemOpen
  \bibfield  {author} {\bibinfo {author} {\bibfnamefont {A.}~\bibnamefont
  {Kaiser}}\ and\ \bibinfo {author} {\bibfnamefont {H.}~\bibnamefont
  {L{\"o}wen}},\ }\bibfield  {title} {\enquote {\bibinfo {title} {Unusual
  swelling of a polymer in a bacterial bath},}\ }\href
  {https://doi.org/http://dx.doi.org/10.1063/1.4891095} {\bibfield  {journal}
  {\bibinfo  {journal} {J. Chem. Phys.}\ }\textbf {\bibinfo {volume} {141}},\
  \bibinfo {eid} {044903} (\bibinfo {year} {2014})}\BibitemShut {NoStop}%
\bibitem [{\citenamefont {Eisenstecken}, \citenamefont {Gompper},\ and\
  \citenamefont {Winkler}(2016)}]{eise:16}%
  \BibitemOpen
  \bibfield  {author} {\bibinfo {author} {\bibfnamefont {T.}~\bibnamefont
  {Eisenstecken}}, \bibinfo {author} {\bibfnamefont {G.}~\bibnamefont
  {Gompper}},\ and\ \bibinfo {author} {\bibfnamefont {R.~G.}\ \bibnamefont
  {Winkler}},\ }\bibfield  {title} {\enquote {\bibinfo {title} {Conformational
  properties of active semiflexible polymers},}\ }\href
  {https://doi.org/10.3390/polym8080304} {\bibfield  {journal} {\bibinfo
  {journal} {Polymers}\ }\textbf {\bibinfo {volume} {8}},\ \bibinfo {pages}
  {304} (\bibinfo {year} {2016})}\BibitemShut {NoStop}%
\bibitem [{\citenamefont {Anand}\ and\ \citenamefont {Singh}(2018)}]{anand:18}%
  \BibitemOpen
  \bibfield  {author} {\bibinfo {author} {\bibfnamefont {S.~K.}\ \bibnamefont
  {Anand}}\ and\ \bibinfo {author} {\bibfnamefont {S.~P.}\ \bibnamefont
  {Singh}},\ }\bibfield  {title} {\enquote {\bibinfo {title} {Structure and
  dynamics of a self-propelled semiflexible filament},}\ }\href@noop {}
  {\bibfield  {journal} {\bibinfo  {journal} {Phys. Rev. E}\ }\textbf {\bibinfo
  {volume} {98}},\ \bibinfo {pages} {042501} (\bibinfo {year}
  {2018})}\BibitemShut {NoStop}%
\bibitem [{\citenamefont {Isele-Holder}, \citenamefont {Elgeti},\ and\
  \citenamefont {Gompper}(2015)}]{isele:15}%
  \BibitemOpen
  \bibfield  {author} {\bibinfo {author} {\bibfnamefont {R.~E.}\ \bibnamefont
  {Isele-Holder}}, \bibinfo {author} {\bibfnamefont {J.}~\bibnamefont
  {Elgeti}},\ and\ \bibinfo {author} {\bibfnamefont {G.}~\bibnamefont
  {Gompper}},\ }\bibfield  {title} {\enquote {\bibinfo {title} {Self-propelled
  worm-like filaments: spontaneous spiral formation, structure, and
  dynamics},}\ }\href@noop {} {\bibfield  {journal} {\bibinfo  {journal} {Soft
  Matter}\ }\textbf {\bibinfo {volume} {11}},\ \bibinfo {pages} {7181}
  (\bibinfo {year} {2015})}\BibitemShut {NoStop}%
\bibitem [{\citenamefont {Bianco}, \citenamefont {Locatelli},\ and\
  \citenamefont {Malgaretti}(2018)}]{bian:18}%
  \BibitemOpen
  \bibfield  {author} {\bibinfo {author} {\bibfnamefont {V.}~\bibnamefont
  {Bianco}}, \bibinfo {author} {\bibfnamefont {E.}~\bibnamefont {Locatelli}},\
  and\ \bibinfo {author} {\bibfnamefont {P.}~\bibnamefont {Malgaretti}},\
  }\bibfield  {title} {\enquote {\bibinfo {title} {Globulelike conformation and
  enhanced diffusion of active polymers},}\ }\href
  {https://doi.org/10.1103/PhysRevLett.121.217802} {\bibfield  {journal}
  {\bibinfo  {journal} {Phys. Rev. Lett.}\ }\textbf {\bibinfo {volume} {121}},\
  \bibinfo {pages} {217802} (\bibinfo {year} {2018})}\BibitemShut {NoStop}%
\bibitem [{\citenamefont {Peterson}, \citenamefont {Hagan},\ and\ \citenamefont
  {Baskaran}(2020)}]{pete:20}%
  \BibitemOpen
  \bibfield  {author} {\bibinfo {author} {\bibfnamefont {M.~S.~E.}\
  \bibnamefont {Peterson}}, \bibinfo {author} {\bibfnamefont {M.~F.}\
  \bibnamefont {Hagan}},\ and\ \bibinfo {author} {\bibfnamefont
  {A.}~\bibnamefont {Baskaran}},\ }\bibfield  {title} {\enquote {\bibinfo
  {title} {Statistical properties of a tangentially driven active filament},}\
  }\href {https://doi.org/10.1088/1742-5468/ab6097} {\bibfield  {journal}
  {\bibinfo  {journal} {J. Stat. Mech. Theor. Exp.}\ }\textbf {\bibinfo
  {volume} {2020}},\ \bibinfo {pages} {013216} (\bibinfo {year}
  {2020})}\BibitemShut {NoStop}%
\bibitem [{\citenamefont {Philipps}, \citenamefont {Gompper},\ and\
  \citenamefont {Winkler}(2022{\natexlab{a}})}]{phil:22.1}%
  \BibitemOpen
  \bibfield  {author} {\bibinfo {author} {\bibfnamefont {C.~A.}\ \bibnamefont
  {Philipps}}, \bibinfo {author} {\bibfnamefont {G.}~\bibnamefont {Gompper}},\
  and\ \bibinfo {author} {\bibfnamefont {R.~G.}\ \bibnamefont {Winkler}},\
  }\bibfield  {title} {\enquote {\bibinfo {title} {Tangentially driven active
  polar linear polymers - an analytical study},}\ }\href@noop {} {\bibfield
  {journal} {\bibinfo  {journal} {J. Chem. Phys.}\ }\textbf {\bibinfo {volume}
  {157}},\ \bibinfo {pages} {194904} (\bibinfo {year}
  {2022}{\natexlab{a}})}\BibitemShut {NoStop}%
\bibitem [{\citenamefont {Philipps}, \citenamefont {Gompper},\ and\
  \citenamefont {Winkler}(2022{\natexlab{b}})}]{phil:22}%
  \BibitemOpen
  \bibfield  {author} {\bibinfo {author} {\bibfnamefont {C.~A.}\ \bibnamefont
  {Philipps}}, \bibinfo {author} {\bibfnamefont {G.}~\bibnamefont {Gompper}},\
  and\ \bibinfo {author} {\bibfnamefont {R.~G.}\ \bibnamefont {Winkler}},\
  }\bibfield  {title} {\enquote {\bibinfo {title} {Dynamics of active polar
  ring polymers},}\ }\href@noop {} {\bibfield  {journal} {\bibinfo  {journal}
  {Phys. Rev. E}\ }\textbf {\bibinfo {volume} {105}},\ \bibinfo {pages}
  {L062501} (\bibinfo {year} {2022}{\natexlab{b}})}\BibitemShut {NoStop}%
\bibitem [{\citenamefont {Fazelzadeh}\ \emph {et~al.}(2023)\citenamefont
  {Fazelzadeh}, \citenamefont {Irani}, \citenamefont {Mokhtari},\ and\
  \citenamefont {Jabbari-Farouji}}]{faze:23}%
  \BibitemOpen
  \bibfield  {author} {\bibinfo {author} {\bibfnamefont {M.}~\bibnamefont
  {Fazelzadeh}}, \bibinfo {author} {\bibfnamefont {E.}~\bibnamefont {Irani}},
  \bibinfo {author} {\bibfnamefont {Z.}~\bibnamefont {Mokhtari}},\ and\
  \bibinfo {author} {\bibfnamefont {S.}~\bibnamefont {Jabbari-Farouji}},\
  }\bibfield  {title} {\enquote {\bibinfo {title} {Effects of inertia on
  conformation and dynamics of tangentially driven active filaments},}\
  }\href@noop {} {\bibfield  {journal} {\bibinfo  {journal} {Phys. Rev. E}\
  }\textbf {\bibinfo {volume} {108}},\ \bibinfo {pages} {024606} (\bibinfo
  {year} {2023})}\BibitemShut {NoStop}%
\bibitem [{\citenamefont {Tejedor}, \citenamefont {Ram{\'\i}rez},\ and\
  \citenamefont {Ripoll}(2024)}]{teje:24}%
  \BibitemOpen
  \bibfield  {author} {\bibinfo {author} {\bibfnamefont {A.~R.}\ \bibnamefont
  {Tejedor}}, \bibinfo {author} {\bibfnamefont {J.}~\bibnamefont
  {Ram{\'\i}rez}},\ and\ \bibinfo {author} {\bibfnamefont {M.}~\bibnamefont
  {Ripoll}},\ }\bibfield  {title} {\enquote {\bibinfo {title} {Progressive
  polymer deformation induced by polar activity and the influence of
  inertia},}\ }\href@noop {} {\bibfield  {journal} {\bibinfo  {journal} {Phys.
  Rev. Res.}\ }\textbf {\bibinfo {volume} {6}},\ \bibinfo {pages} {L032002}
  (\bibinfo {year} {2024})}\BibitemShut {NoStop}%
\bibitem [{\citenamefont {Winkler}(2025)}]{wink:25}%
  \BibitemOpen
  \bibfield  {author} {\bibinfo {author} {\bibfnamefont {R.~G.}\ \bibnamefont
  {Winkler}},\ }\bibfield  {title} {\enquote {\bibinfo {title} {Conformational
  properties of active polar semiflexible phantom polymers},}\ }\href@noop {}
  {\bibfield  {journal} {\bibinfo  {journal} {J. Chem. Phys.}\ }\textbf
  {\bibinfo {volume} {162}},\ \bibinfo {pages} {154903} (\bibinfo {year}
  {2025})}\BibitemShut {NoStop}%
\bibitem [{\citenamefont {Karan}, \citenamefont {Chaudhuri},\ and\
  \citenamefont {Chaudhuri}(2024)}]{kara:24}%
  \BibitemOpen
  \bibfield  {author} {\bibinfo {author} {\bibfnamefont {C.}~\bibnamefont
  {Karan}}, \bibinfo {author} {\bibfnamefont {A.}~\bibnamefont {Chaudhuri}},\
  and\ \bibinfo {author} {\bibfnamefont {D.}~\bibnamefont {Chaudhuri}},\
  }\bibfield  {title} {\enquote {\bibinfo {title} {Inertia and activity: spiral
  transitions in semi-flexible, self-avoiding polymers},}\ }\href@noop {}
  {\bibfield  {journal} {\bibinfo  {journal} {Soft Matter}\ }\textbf {\bibinfo
  {volume} {20}},\ \bibinfo {pages} {6221--6230} (\bibinfo {year}
  {2024})}\BibitemShut {NoStop}%
\bibitem [{\citenamefont {Sanchez}\ \emph {et~al.}(2012)\citenamefont
  {Sanchez}, \citenamefont {Chen}, \citenamefont {DeCamp}, \citenamefont
  {Heymann},\ and\ \citenamefont {Dogic}}]{sanc:12}%
  \BibitemOpen
  \bibfield  {author} {\bibinfo {author} {\bibfnamefont {T.}~\bibnamefont
  {Sanchez}}, \bibinfo {author} {\bibfnamefont {D.~T.~N.}\ \bibnamefont
  {Chen}}, \bibinfo {author} {\bibfnamefont {S.~J.}\ \bibnamefont {DeCamp}},
  \bibinfo {author} {\bibfnamefont {M.}~\bibnamefont {Heymann}},\ and\ \bibinfo
  {author} {\bibfnamefont {Z.}~\bibnamefont {Dogic}},\ }\bibfield  {title}
  {\enquote {\bibinfo {title} {Spontaneous motion in hierarchically assembled
  active matter},}\ }\href {https://doi.org/10.1038/nature11591} {\bibfield
  {journal} {\bibinfo  {journal} {Nature}\ }\textbf {\bibinfo {volume} {491}},\
  \bibinfo {pages} {431} (\bibinfo {year} {2012})}\BibitemShut {NoStop}%
\bibitem [{\citenamefont {Doostmohammadi}\ \emph {et~al.}(2018)\citenamefont
  {Doostmohammadi}, \citenamefont {Ign{\'e}s-Mullol}, \citenamefont {Yeomans},\
  and\ \citenamefont {Sagu{\'e}s}}]{doos:18}%
  \BibitemOpen
  \bibfield  {author} {\bibinfo {author} {\bibfnamefont {A.}~\bibnamefont
  {Doostmohammadi}}, \bibinfo {author} {\bibfnamefont {J.}~\bibnamefont
  {Ign{\'e}s-Mullol}}, \bibinfo {author} {\bibfnamefont {J.~M.}\ \bibnamefont
  {Yeomans}},\ and\ \bibinfo {author} {\bibfnamefont {F.}~\bibnamefont
  {Sagu{\'e}s}},\ }\bibfield  {title} {\enquote {\bibinfo {title} {Active
  nematics},}\ }\href {https://doi.org/10.1038/s41467-018-05666-8} {\bibfield
  {journal} {\bibinfo  {journal} {Nat. Commun.}\ }\textbf {\bibinfo {volume}
  {9}},\ \bibinfo {pages} {3246} (\bibinfo {year} {2018})}\BibitemShut
  {NoStop}%
\bibitem [{\citenamefont {Schaller}\ \emph {et~al.}(2010)\citenamefont
  {Schaller}, \citenamefont {Weber}, \citenamefont {Semmrich}, \citenamefont
  {Frey},\ and\ \citenamefont {Bausch}}]{scha:10}%
  \BibitemOpen
  \bibfield  {author} {\bibinfo {author} {\bibfnamefont {V.}~\bibnamefont
  {Schaller}}, \bibinfo {author} {\bibfnamefont {C.}~\bibnamefont {Weber}},
  \bibinfo {author} {\bibfnamefont {C.}~\bibnamefont {Semmrich}}, \bibinfo
  {author} {\bibfnamefont {E.}~\bibnamefont {Frey}},\ and\ \bibinfo {author}
  {\bibfnamefont {A.~R.}\ \bibnamefont {Bausch}},\ }\bibfield  {title}
  {\enquote {\bibinfo {title} {Polar patterns of driven filaments},}\ }\href
  {https://doi.org/10.1038/nature09312} {\bibfield  {journal} {\bibinfo
  {journal} {Nature}\ }\textbf {\bibinfo {volume} {467}},\ \bibinfo {pages}
  {73} (\bibinfo {year} {2010})}\BibitemShut {NoStop}%
\bibitem [{\citenamefont {Kawamura}\ \emph {et~al.}(2008)\citenamefont
  {Kawamura}, \citenamefont {Kakugo}, \citenamefont {Shikinaka}, \citenamefont
  {Osada},\ and\ \citenamefont {Gong}}]{kawa:08}%
  \BibitemOpen
  \bibfield  {author} {\bibinfo {author} {\bibfnamefont {R.}~\bibnamefont
  {Kawamura}}, \bibinfo {author} {\bibfnamefont {A.}~\bibnamefont {Kakugo}},
  \bibinfo {author} {\bibfnamefont {K.}~\bibnamefont {Shikinaka}}, \bibinfo
  {author} {\bibfnamefont {Y.}~\bibnamefont {Osada}},\ and\ \bibinfo {author}
  {\bibfnamefont {J.~P.}\ \bibnamefont {Gong}},\ }\bibfield  {title} {\enquote
  {\bibinfo {title} {Ring-shaped assembly of microtubules shows preferential
  counterclockwise motion},}\ }\href@noop {} {\bibfield  {journal} {\bibinfo
  {journal} {Biomacromolecules}\ }\textbf {\bibinfo {volume} {9}},\ \bibinfo
  {pages} {2277} (\bibinfo {year} {2008})}\BibitemShut {NoStop}%
\bibitem [{\citenamefont {Liu}, \citenamefont {T{\"u}zel},\ and\ \citenamefont
  {Ross}(2011)}]{liu:11}%
  \BibitemOpen
  \bibfield  {author} {\bibinfo {author} {\bibfnamefont {L.}~\bibnamefont
  {Liu}}, \bibinfo {author} {\bibfnamefont {E.}~\bibnamefont {T{\"u}zel}},\
  and\ \bibinfo {author} {\bibfnamefont {J.~L.}\ \bibnamefont {Ross}},\
  }\bibfield  {title} {\enquote {\bibinfo {title} {Loop formation of
  microtubules during gliding at high density},}\ }\href@noop {} {\bibfield
  {journal} {\bibinfo  {journal} {J. Phys.: Condens. Matter}\ }\textbf
  {\bibinfo {volume} {23}},\ \bibinfo {pages} {374104} (\bibinfo {year}
  {2011})}\BibitemShut {NoStop}%
\bibitem [{\citenamefont {Keya}, \citenamefont {Kabir},\ and\ \citenamefont
  {Kakugo}(2020)}]{keya:20}%
  \BibitemOpen
  \bibfield  {author} {\bibinfo {author} {\bibfnamefont {J.~J.}\ \bibnamefont
  {Keya}}, \bibinfo {author} {\bibfnamefont {A.~M.~R.}\ \bibnamefont {Kabir}},\
  and\ \bibinfo {author} {\bibfnamefont {A.}~\bibnamefont {Kakugo}},\
  }\bibfield  {title} {\enquote {\bibinfo {title} {Synchronous operation of
  biomolecular engines},}\ }\href@noop {} {\bibfield  {journal} {\bibinfo
  {journal} {Biophys. Rev.}\ }\textbf {\bibinfo {volume} {12}},\ \bibinfo
  {pages} {401} (\bibinfo {year} {2020})}\BibitemShut {NoStop}%
\bibitem [{\citenamefont {Locatelli}, \citenamefont {Bianco},\ and\
  \citenamefont {Malgaretti}(2021)}]{loca:21}%
  \BibitemOpen
  \bibfield  {author} {\bibinfo {author} {\bibfnamefont {E.}~\bibnamefont
  {Locatelli}}, \bibinfo {author} {\bibfnamefont {V.}~\bibnamefont {Bianco}},\
  and\ \bibinfo {author} {\bibfnamefont {P.}~\bibnamefont {Malgaretti}},\
  }\bibfield  {title} {\enquote {\bibinfo {title} {Active polymer rings:
  activity-induced collapse and dynamicl arrest},}\ }\href@noop {} {\bibfield
  {journal} {\bibinfo  {journal} {Phys. Rev. Lett.}\ }\textbf {\bibinfo
  {volume} {126}},\ \bibinfo {pages} {097801} (\bibinfo {year}
  {2021})}\BibitemShut {NoStop}%
\bibitem [{\citenamefont {Lamura}(2024{\natexlab{a}})}]{lamu:24_2}%
  \BibitemOpen
  \bibfield  {author} {\bibinfo {author} {\bibfnamefont {A.}~\bibnamefont
  {Lamura}},\ }\bibfield  {title} {\enquote {\bibinfo {title} {Excluded volume
  effects on tangentially driven active ring polymers},}\ }\href
  {https://doi.org/10.1103/PhysRevE.109.054611} {\bibfield  {journal} {\bibinfo
   {journal} {Phys. Rev. E}\ }\textbf {\bibinfo {volume} {109}},\ \bibinfo
  {pages} {054611} (\bibinfo {year} {2024}{\natexlab{a}})}\BibitemShut
  {NoStop}%
\bibitem [{\citenamefont {Janzen}\ and\ \citenamefont
  {Matoz-Fernandez}(2024)}]{janz:24}%
  \BibitemOpen
  \bibfield  {author} {\bibinfo {author} {\bibfnamefont {G.}~\bibnamefont
  {Janzen}}\ and\ \bibinfo {author} {\bibfnamefont {D.~A.}\ \bibnamefont
  {Matoz-Fernandez}},\ }\bibfield  {title} {\enquote {\bibinfo {title} {Density
  and inertia effects on two-dimensional active semiflexible filament
  suspensions},}\ }\href {https://doi.org/10.1039/D4SM00572D} {\bibfield
  {journal} {\bibinfo  {journal} {Soft Matter}\ }\textbf {\bibinfo {volume}
  {20}},\ \bibinfo {pages} {6618} (\bibinfo {year} {2024})}\BibitemShut
  {NoStop}%
\bibitem [{\citenamefont {Huang}\ \emph {et~al.}(2010)\citenamefont {Huang},
  \citenamefont {Winkler}, \citenamefont {Sutmann},\ and\ \citenamefont
  {Gompper}}]{huan:10}%
  \BibitemOpen
  \bibfield  {author} {\bibinfo {author} {\bibfnamefont {C.-C.}\ \bibnamefont
  {Huang}}, \bibinfo {author} {\bibfnamefont {R.~G.}\ \bibnamefont {Winkler}},
  \bibinfo {author} {\bibfnamefont {G.}~\bibnamefont {Sutmann}},\ and\ \bibinfo
  {author} {\bibfnamefont {G.}~\bibnamefont {Gompper}},\ }\bibfield  {title}
  {\enquote {\bibinfo {title} {Semidilute polymer solutions at equilibrium and
  under shear flow},}\ }\href@noop {} {\bibfield  {journal} {\bibinfo
  {journal} {Macromolecules}\ }\textbf {\bibinfo {volume} {43}},\ \bibinfo
  {pages} {10107} (\bibinfo {year} {2010})}\BibitemShut {NoStop}%
\bibitem [{\citenamefont {Winkler}(2010)}]{wink:10}%
  \BibitemOpen
  \bibfield  {author} {\bibinfo {author} {\bibfnamefont {R.~G.}\ \bibnamefont
  {Winkler}},\ }\bibfield  {title} {\enquote {\bibinfo {title} {Conformational
  and rheological properties of semiflexible polymers in shear flow},}\
  }\href@noop {} {\bibfield  {journal} {\bibinfo  {journal} {J. Chem. Phys.}\
  }\textbf {\bibinfo {volume} {133}},\ \bibinfo {pages} {164905} (\bibinfo
  {year} {2010})}\BibitemShut {NoStop}%
\bibitem [{\citenamefont {Schroeder}\ \emph
  {et~al.}(2005{\natexlab{a}})\citenamefont {Schroeder}, \citenamefont
  {Teixeira}, \citenamefont {Shaqfeh},\ and\ \citenamefont {Chu}}]{schr:05.1}%
  \BibitemOpen
  \bibfield  {author} {\bibinfo {author} {\bibfnamefont {C.~M.}\ \bibnamefont
  {Schroeder}}, \bibinfo {author} {\bibfnamefont {R.~E.}\ \bibnamefont
  {Teixeira}}, \bibinfo {author} {\bibfnamefont {E.~S.~G.}\ \bibnamefont
  {Shaqfeh}},\ and\ \bibinfo {author} {\bibfnamefont {S.}~\bibnamefont {Chu}},\
  }\bibfield  {title} {\enquote {\bibinfo {title} {Dynamics of {DNA} in the
  flow-gradient plane of steady shear flow: Observations and simulations},}\
  }\href@noop {} {\bibfield  {journal} {\bibinfo  {journal} {Macromolecules}\
  }\textbf {\bibinfo {volume} {38}},\ \bibinfo {pages} {1967} (\bibinfo {year}
  {2005}{\natexlab{a}})}\BibitemShut {NoStop}%
\bibitem [{\citenamefont {Lamura}\ and\ \citenamefont
  {Winkler}(2012)}]{lamu:12}%
  \BibitemOpen
  \bibfield  {author} {\bibinfo {author} {\bibfnamefont {A.}~\bibnamefont
  {Lamura}}\ and\ \bibinfo {author} {\bibfnamefont {R.~G.}\ \bibnamefont
  {Winkler}},\ }\bibfield  {title} {\enquote {\bibinfo {title} {Semiflexible
  polymers under external fields confined to two dimensions},}\ }\href@noop {}
  {\bibfield  {journal} {\bibinfo  {journal} {J. Chem. Phys.}\ }\textbf
  {\bibinfo {volume} {137}},\ \bibinfo {pages} {244909} (\bibinfo {year}
  {2012})}\BibitemShut {NoStop}%
\bibitem [{\citenamefont {Smith}, \citenamefont {Babcock},\ and\ \citenamefont
  {Chu}(1999)}]{smit:99}%
  \BibitemOpen
  \bibfield  {author} {\bibinfo {author} {\bibfnamefont {D.~E.}\ \bibnamefont
  {Smith}}, \bibinfo {author} {\bibfnamefont {H.~P.}\ \bibnamefont {Babcock}},\
  and\ \bibinfo {author} {\bibfnamefont {S.}~\bibnamefont {Chu}},\ }\bibfield
  {title} {\enquote {\bibinfo {title} {Single polymer dynamics in steady shear
  flow},}\ }\href@noop {} {\bibfield  {journal} {\bibinfo  {journal} {Science}\
  }\textbf {\bibinfo {volume} {283}},\ \bibinfo {pages} {1724} (\bibinfo {year}
  {1999})}\BibitemShut {NoStop}%
\bibitem [{\citenamefont {Teixeira}\ \emph {et~al.}(2005)\citenamefont
  {Teixeira}, \citenamefont {Babcock}, \citenamefont {Shaqfeh},\ and\
  \citenamefont {Chu}}]{teix:05}%
  \BibitemOpen
  \bibfield  {author} {\bibinfo {author} {\bibfnamefont {R.~E.}\ \bibnamefont
  {Teixeira}}, \bibinfo {author} {\bibfnamefont {H.~P.}\ \bibnamefont
  {Babcock}}, \bibinfo {author} {\bibfnamefont {E.~S.~G.}\ \bibnamefont
  {Shaqfeh}},\ and\ \bibinfo {author} {\bibfnamefont {S.}~\bibnamefont {Chu}},\
  }\bibfield  {title} {\enquote {\bibinfo {title} {Shear thinning and tumbling
  dynamics of single polymers in the flow-gradient plane},}\ }\href@noop {}
  {\bibfield  {journal} {\bibinfo  {journal} {Macromolecules}\ }\textbf
  {\bibinfo {volume} {38}},\ \bibinfo {pages} {581} (\bibinfo {year}
  {2005})}\BibitemShut {NoStop}%
\bibitem [{\citenamefont {Winkler}, \citenamefont {Keller},\ and\ \citenamefont
  {R{\"a}dler}(2006)}]{wink:06}%
  \BibitemOpen
  \bibfield  {author} {\bibinfo {author} {\bibfnamefont {R.~G.}\ \bibnamefont
  {Winkler}}, \bibinfo {author} {\bibfnamefont {S.}~\bibnamefont {Keller}},\
  and\ \bibinfo {author} {\bibfnamefont {J.~O.}\ \bibnamefont {R{\"a}dler}},\
  }\bibfield  {title} {\enquote {\bibinfo {title} {Intramolecular dynamics of
  linear macromolecules by fluorescence correlation spectroscopy},}\
  }\href@noop {} {\bibfield  {journal} {\bibinfo  {journal} {Phys. Rev. E}\
  }\textbf {\bibinfo {volume} {73}},\ \bibinfo {pages} {041919} (\bibinfo
  {year} {2006})}\BibitemShut {NoStop}%
\bibitem [{\citenamefont {Schroeder}\ \emph
  {et~al.}(2005{\natexlab{b}})\citenamefont {Schroeder}, \citenamefont
  {Teixeira}, \citenamefont {Shaqfeh},\ and\ \citenamefont {Chu}}]{schr:05}%
  \BibitemOpen
  \bibfield  {author} {\bibinfo {author} {\bibfnamefont {C.~M.}\ \bibnamefont
  {Schroeder}}, \bibinfo {author} {\bibfnamefont {R.~E.}\ \bibnamefont
  {Teixeira}}, \bibinfo {author} {\bibfnamefont {E.~S.~G.}\ \bibnamefont
  {Shaqfeh}},\ and\ \bibinfo {author} {\bibfnamefont {S.}~\bibnamefont {Chu}},\
  }\bibfield  {title} {\enquote {\bibinfo {title} {Characteristic periodic
  motion of polymers in shear flow},}\ }\href@noop {} {\bibfield  {journal}
  {\bibinfo  {journal} {Phys. Rev. Lett.}\ }\textbf {\bibinfo {volume} {95}},\
  \bibinfo {pages} {018301} (\bibinfo {year} {2005}{\natexlab{b}})}\BibitemShut
  {NoStop}%
\bibitem [{\citenamefont {Puliafito}\ and\ \citenamefont
  {Turitsyn}(2005)}]{puli:05}%
  \BibitemOpen
  \bibfield  {author} {\bibinfo {author} {\bibfnamefont {A.}~\bibnamefont
  {Puliafito}}\ and\ \bibinfo {author} {\bibfnamefont {K.}~\bibnamefont
  {Turitsyn}},\ }\bibfield  {title} {\enquote {\bibinfo {title} {Numerical
  study of polymer tumbling in linear shear flow},}\ }\href@noop {} {\bibfield
  {journal} {\bibinfo  {journal} {Physica D}\ }\textbf {\bibinfo {volume}
  {211}},\ \bibinfo {pages} {9} (\bibinfo {year} {2005})}\BibitemShut {NoStop}%
\bibitem [{\citenamefont {Pincus}, \citenamefont {Rodger},\ and\ \citenamefont
  {Ravi~Prakash}(2023)}]{pinc:23}%
  \BibitemOpen
  \bibfield  {author} {\bibinfo {author} {\bibfnamefont {I.}~\bibnamefont
  {Pincus}}, \bibinfo {author} {\bibfnamefont {A.}~\bibnamefont {Rodger}},\
  and\ \bibinfo {author} {\bibfnamefont {J.}~\bibnamefont {Ravi~Prakash}},\
  }\bibfield  {title} {\enquote {\bibinfo {title} {Dilute polymer solutions
  under shear flow: Comprehensive qualitative analysis using a bead-spring
  chain model with a {FENE-F}raenkel spring},}\ }\href@noop {} {\bibfield
  {journal} {\bibinfo  {journal} {Journal of Rheology}\ }\textbf {\bibinfo
  {volume} {67}},\ \bibinfo {pages} {373--402} (\bibinfo {year}
  {2023})}\BibitemShut {NoStop}%
\bibitem [{\citenamefont {Hatwalne}\ \emph {et~al.}(2004)\citenamefont
  {Hatwalne}, \citenamefont {Ramaswamy}, \citenamefont {Rao},\ and\
  \citenamefont {Simha}}]{hatw:04}%
  \BibitemOpen
  \bibfield  {author} {\bibinfo {author} {\bibfnamefont {Y.}~\bibnamefont
  {Hatwalne}}, \bibinfo {author} {\bibfnamefont {S.}~\bibnamefont {Ramaswamy}},
  \bibinfo {author} {\bibfnamefont {M.}~\bibnamefont {Rao}},\ and\ \bibinfo
  {author} {\bibfnamefont {R.~A.}\ \bibnamefont {Simha}},\ }\bibfield  {title}
  {\enquote {\bibinfo {title} {Rheology of active-particle suspensions},}\
  }\href {https://doi.org/10.1103/PhysRevLett.92.118101} {\bibfield  {journal}
  {\bibinfo  {journal} {Phys. Rev. Lett.}\ }\textbf {\bibinfo {volume} {92}},\
  \bibinfo {pages} {118101} (\bibinfo {year} {2004})}\BibitemShut {NoStop}%
\bibitem [{\citenamefont {Saintillan}(2010)}]{sain:10}%
  \BibitemOpen
  \bibfield  {author} {\bibinfo {author} {\bibfnamefont {D.}~\bibnamefont
  {Saintillan}},\ }\bibfield  {title} {\enquote {\bibinfo {title} {The dilute
  rheology of swimming suspensions: A simple kinetic model},}\ }\href
  {https://doi.org/10.1007/s11340-009-9267-0} {\bibfield  {journal} {\bibinfo
  {journal} {Exp. Mech.}\ }\textbf {\bibinfo {volume} {50}},\ \bibinfo {pages}
  {1275} (\bibinfo {year} {2010})}\BibitemShut {NoStop}%
\bibitem [{\citenamefont {Marchetti}\ \emph {et~al.}(2013)\citenamefont
  {Marchetti}, \citenamefont {Joanny}, \citenamefont {Ramaswamy}, \citenamefont
  {Liverpool}, \citenamefont {Prost}, \citenamefont {Rao},\ and\ \citenamefont
  {Simha}}]{marc:13}%
  \BibitemOpen
  \bibfield  {author} {\bibinfo {author} {\bibfnamefont {M.~C.}\ \bibnamefont
  {Marchetti}}, \bibinfo {author} {\bibfnamefont {J.~F.}\ \bibnamefont
  {Joanny}}, \bibinfo {author} {\bibfnamefont {S.}~\bibnamefont {Ramaswamy}},
  \bibinfo {author} {\bibfnamefont {T.~B.}\ \bibnamefont {Liverpool}}, \bibinfo
  {author} {\bibfnamefont {J.}~\bibnamefont {Prost}}, \bibinfo {author}
  {\bibfnamefont {M.}~\bibnamefont {Rao}},\ and\ \bibinfo {author}
  {\bibfnamefont {R.~A.}\ \bibnamefont {Simha}},\ }\bibfield  {title} {\enquote
  {\bibinfo {title} {Hydrodynamics of soft active matter},}\ }\href@noop {}
  {\bibfield  {journal} {\bibinfo  {journal} {Rev. Mod. Phys.}\ }\textbf
  {\bibinfo {volume} {85}},\ \bibinfo {pages} {1143} (\bibinfo {year}
  {2013})}\BibitemShut {NoStop}%
\bibitem [{\citenamefont {Lamura}\ and\ \citenamefont
  {Winkler}(2019)}]{lamu:19}%
  \BibitemOpen
  \bibfield  {author} {\bibinfo {author} {\bibfnamefont {A.}~\bibnamefont
  {Lamura}}\ and\ \bibinfo {author} {\bibfnamefont {R.~G.}\ \bibnamefont
  {Winkler}},\ }\bibfield  {title} {\enquote {\bibinfo {title} {Tethered
  semiflexible polymer under large amplitude oscillatory shear},}\ }\href@noop
  {} {\bibfield  {journal} {\bibinfo  {journal} {Polymers}\ }\textbf {\bibinfo
  {volume} {11}},\ \bibinfo {pages} {737} (\bibinfo {year} {2019})}\BibitemShut
  {NoStop}%
\bibitem [{\citenamefont {Lamura}, \citenamefont {Winkler},\ and\ \citenamefont
  {Gompper}(2021)}]{lamu:21}%
  \BibitemOpen
  \bibfield  {author} {\bibinfo {author} {\bibfnamefont {A.}~\bibnamefont
  {Lamura}}, \bibinfo {author} {\bibfnamefont {R.~G.}\ \bibnamefont
  {Winkler}},\ and\ \bibinfo {author} {\bibfnamefont {G.}~\bibnamefont
  {Gompper}},\ }\bibfield  {title} {\enquote {\bibinfo {title} {Wall-anchored
  semiflexible polymer under large amplitude oscillatory shear flow},}\
  }\href@noop {} {\bibfield  {journal} {\bibinfo  {journal} {J. Chem. Phys.}\
  }\textbf {\bibinfo {volume} {154}},\ \bibinfo {pages} {224901} (\bibinfo
  {year} {2021})}\BibitemShut {NoStop}%
\bibitem [{\citenamefont {Allen}\ and\ \citenamefont
  {Tildesley}(1987)}]{alle:87}%
  \BibitemOpen
  \bibfield  {author} {\bibinfo {author} {\bibfnamefont {M.~P.}\ \bibnamefont
  {Allen}}\ and\ \bibinfo {author} {\bibfnamefont {D.~J.}\ \bibnamefont
  {Tildesley}},\ }\href@noop {} {\emph {\bibinfo {title} {Computer Simulation
  of Liquids}}}\ (\bibinfo  {publisher} {Clarendon Press},\ \bibinfo {address}
  {Oxford},\ \bibinfo {year} {1987})\BibitemShut {NoStop}%
\bibitem [{\citenamefont {Jiang}\ and\ \citenamefont {Hou}(2014)}]{jiang:14}%
  \BibitemOpen
  \bibfield  {author} {\bibinfo {author} {\bibfnamefont {H.}~\bibnamefont
  {Jiang}}\ and\ \bibinfo {author} {\bibfnamefont {Z.}~\bibnamefont {Hou}},\
  }\bibfield  {title} {\enquote {\bibinfo {title} {Motion transition of active
  filaments: rotation without hydrodynamic interactions},}\ }\href@noop {}
  {\bibfield  {journal} {\bibinfo  {journal} {Soft Matter}\ }\textbf {\bibinfo
  {volume} {10}},\ \bibinfo {pages} {1012} (\bibinfo {year}
  {2014})}\BibitemShut {NoStop}%
\bibitem [{\citenamefont {Philipps}, \citenamefont {Gompper},\ and\
  \citenamefont {Winkler}(2022{\natexlab{c}})}]{phil:22_1}%
  \BibitemOpen
  \bibfield  {author} {\bibinfo {author} {\bibfnamefont {C.~A.}\ \bibnamefont
  {Philipps}}, \bibinfo {author} {\bibfnamefont {G.}~\bibnamefont {Gompper}},\
  and\ \bibinfo {author} {\bibfnamefont {R.~G.}\ \bibnamefont {Winkler}},\
  }\bibfield  {title} {\enquote {\bibinfo {title} {Tangentially driven active
  polar linear polymers - an analytical study},}\ }\href@noop {} {\bibfield
  {journal} {\bibinfo  {journal} {J. Chem. Phys.}\ }\textbf {\bibinfo {volume}
  {157}},\ \bibinfo {pages} {194904} (\bibinfo {year}
  {2022}{\natexlab{c}})}\BibitemShut {NoStop}%
\bibitem [{\citenamefont {Janzen}\ \emph {et~al.}(2025)\citenamefont {Janzen},
  \citenamefont {Miranda}, \citenamefont {Mart{\'\i}n-Roca}, \citenamefont
  {Malgaretti}, \citenamefont {Locatelli}, \citenamefont {Valeriani},\ and\
  \citenamefont {Fernandez}}]{janz:25}%
  \BibitemOpen
  \bibfield  {author} {\bibinfo {author} {\bibfnamefont {G.}~\bibnamefont
  {Janzen}}, \bibinfo {author} {\bibfnamefont {J.~P.}\ \bibnamefont {Miranda}},
  \bibinfo {author} {\bibfnamefont {J.}~\bibnamefont {Mart{\'\i}n-Roca}},
  \bibinfo {author} {\bibfnamefont {P.}~\bibnamefont {Malgaretti}}, \bibinfo
  {author} {\bibfnamefont {E.}~\bibnamefont {Locatelli}}, \bibinfo {author}
  {\bibfnamefont {C.}~\bibnamefont {Valeriani}},\ and\ \bibinfo {author}
  {\bibfnamefont {D.~A.~M.}\ \bibnamefont {Fernandez}},\ }\bibfield  {title}
  {\enquote {\bibinfo {title} {Active polymer behavior in two dimensions: A
  comparative analysis of tangential and push--pull models},}\ }\href
  {https://doi.org/10.1063/5.0243432} {\bibfield  {journal} {\bibinfo
  {journal} {J. Chem. Phys.}\ }\textbf {\bibinfo {volume} {162}},\ \bibinfo
  {pages} {114905} (\bibinfo {year} {2025})}\BibitemShut {NoStop}%
\bibitem [{\citenamefont {Kikuchi}\ \emph {et~al.}(2003)\citenamefont
  {Kikuchi}, \citenamefont {Pooley}, \citenamefont {Ryder},\ and\ \citenamefont
  {Yeomans}}]{kiku:03}%
  \BibitemOpen
  \bibfield  {author} {\bibinfo {author} {\bibfnamefont {N.}~\bibnamefont
  {Kikuchi}}, \bibinfo {author} {\bibfnamefont {C.~M.}\ \bibnamefont {Pooley}},
  \bibinfo {author} {\bibfnamefont {J.~F.}\ \bibnamefont {Ryder}},\ and\
  \bibinfo {author} {\bibfnamefont {J.~M.}\ \bibnamefont {Yeomans}},\
  }\bibfield  {title} {\enquote {\bibinfo {title} {Transport coefficients of a
  mesoscopic fluid dynamics model},}\ }\href@noop {} {\bibfield  {journal}
  {\bibinfo  {journal} {J. Chem. Phys.}\ }\textbf {\bibinfo {volume} {119}},\
  \bibinfo {pages} {6388--6395} (\bibinfo {year} {2003})}\BibitemShut {NoStop}%
\bibitem [{\citenamefont {Ripoll}, \citenamefont {Winkler},\ and\ \citenamefont
  {Gompper}(2007)}]{ripo:07}%
  \BibitemOpen
  \bibfield  {author} {\bibinfo {author} {\bibfnamefont {M.}~\bibnamefont
  {Ripoll}}, \bibinfo {author} {\bibfnamefont {R.~G.}\ \bibnamefont
  {Winkler}},\ and\ \bibinfo {author} {\bibfnamefont {G.}~\bibnamefont
  {Gompper}},\ }\bibfield  {title} {\enquote {\bibinfo {title} {Hydrodynamic
  screening of star polymers in shear flow},}\ }\href@noop {} {\bibfield
  {journal} {\bibinfo  {journal} {Eur. Phys. J. E}\ }\textbf {\bibinfo {volume}
  {23}},\ \bibinfo {pages} {349} (\bibinfo {year} {2007})}\BibitemShut
  {NoStop}%
\bibitem [{\citenamefont {Gompper}\ \emph {et~al.}(2009)\citenamefont
  {Gompper}, \citenamefont {Ihle}, \citenamefont {Kroll},\ and\ \citenamefont
  {Winkler}}]{gomp:09}%
  \BibitemOpen
  \bibfield  {author} {\bibinfo {author} {\bibfnamefont {G.}~\bibnamefont
  {Gompper}}, \bibinfo {author} {\bibfnamefont {T.}~\bibnamefont {Ihle}},
  \bibinfo {author} {\bibfnamefont {D.~M.}\ \bibnamefont {Kroll}},\ and\
  \bibinfo {author} {\bibfnamefont {R.~G.}\ \bibnamefont {Winkler}},\
  }\bibfield  {title} {\enquote {\bibinfo {title} {Multi-particle collision
  dynamics: A particle-based mesoscale simulation approach to the hydrodynamics
  of complex fluids},}\ }\href@noop {} {\bibfield  {journal} {\bibinfo
  {journal} {Adv. Polym. Sci.}\ }\textbf {\bibinfo {volume} {221}},\ \bibinfo
  {pages} {1} (\bibinfo {year} {2009})}\BibitemShut {NoStop}%
\bibitem [{\citenamefont {Lamura}(2024{\natexlab{b}})}]{lamu:24}%
  \BibitemOpen
  \bibfield  {author} {\bibinfo {author} {\bibfnamefont {A.}~\bibnamefont
  {Lamura}},\ }\bibfield  {title} {\enquote {\bibinfo {title} {Tethered
  flexible polymer under oscillatory linear flow},}\ }\href
  {https://doi.org/https://doi.org/10.1016/j.apnum.2024.07.009} {\bibfield
  {journal} {\bibinfo  {journal} {Applied Numerical Mathematics}\ }\textbf
  {\bibinfo {volume} {205}},\ \bibinfo {pages} {206--214} (\bibinfo {year}
  {2024}{\natexlab{b}})}\BibitemShut {NoStop}%
\bibitem [{\citenamefont {LeDuc}\ \emph {et~al.}(1999)\citenamefont {LeDuc},
  \citenamefont {Haber}, \citenamefont {Boa},\ and\ \citenamefont
  {Wirtz}}]{ledu:99}%
  \BibitemOpen
  \bibfield  {author} {\bibinfo {author} {\bibfnamefont {P.}~\bibnamefont
  {LeDuc}}, \bibinfo {author} {\bibfnamefont {C.}~\bibnamefont {Haber}},
  \bibinfo {author} {\bibfnamefont {G.}~\bibnamefont {Boa}},\ and\ \bibinfo
  {author} {\bibfnamefont {D.}~\bibnamefont {Wirtz}},\ }\bibfield  {title}
  {\enquote {\bibinfo {title} {Dynamics of individual flexible polymers in a
  shear flow},}\ }\href@noop {} {\bibfield  {journal} {\bibinfo  {journal}
  {Nature}\ }\textbf {\bibinfo {volume} {399}},\ \bibinfo {pages} {564}
  (\bibinfo {year} {1999})}\BibitemShut {NoStop}%
\bibitem [{\citenamefont {Ryder}\ and\ \citenamefont
  {Yeomans}(2006)}]{ryde:06}%
  \BibitemOpen
  \bibfield  {author} {\bibinfo {author} {\bibfnamefont {J.~F.}\ \bibnamefont
  {Ryder}}\ and\ \bibinfo {author} {\bibfnamefont {J.~M.}\ \bibnamefont
  {Yeomans}},\ }\bibfield  {title} {\enquote {\bibinfo {title} {Shear thinning
  in dilute polymer solutions},}\ }\href@noop {} {\bibfield  {journal}
  {\bibinfo  {journal} {J. Chem. Phys.}\ }\textbf {\bibinfo {volume} {125}},\
  \bibinfo {pages} {194906} (\bibinfo {year} {2006})}\BibitemShut {NoStop}%
\bibitem [{\citenamefont {Gerashchenko}\ and\ \citenamefont
  {Steinberg}(2006)}]{gera:06}%
  \BibitemOpen
  \bibfield  {author} {\bibinfo {author} {\bibfnamefont {S.}~\bibnamefont
  {Gerashchenko}}\ and\ \bibinfo {author} {\bibfnamefont {V.}~\bibnamefont
  {Steinberg}},\ }\bibfield  {title} {\enquote {\bibinfo {title} {Statistics of
  tumbling of a single polymer molecule in shear flow},}\ }\href@noop {}
  {\bibfield  {journal} {\bibinfo  {journal} {Phys. Rev. Lett.}\ }\textbf
  {\bibinfo {volume} {96}},\ \bibinfo {pages} {038304} (\bibinfo {year}
  {2006})}\BibitemShut {NoStop}%
\bibitem [{\citenamefont {Huang}\ \emph {et~al.}(2011)\citenamefont {Huang},
  \citenamefont {Sutmann}, \citenamefont {Gompper},\ and\ \citenamefont
  {Winkler}}]{huan:11}%
  \BibitemOpen
  \bibfield  {author} {\bibinfo {author} {\bibfnamefont {C.-C.}\ \bibnamefont
  {Huang}}, \bibinfo {author} {\bibfnamefont {G.}~\bibnamefont {Sutmann}},
  \bibinfo {author} {\bibfnamefont {G.}~\bibnamefont {Gompper}},\ and\ \bibinfo
  {author} {\bibfnamefont {R.~G.}\ \bibnamefont {Winkler}},\ }\bibfield
  {title} {\enquote {\bibinfo {title} {Tumbling of polymers in semidilute
  solution under shear flow},}\ }\href@noop {} {\bibfield  {journal} {\bibinfo
  {journal} {EPL}\ }\textbf {\bibinfo {volume} {93}},\ \bibinfo {pages} {54004}
  (\bibinfo {year} {2011})}\BibitemShut {NoStop}%
\bibitem [{\citenamefont {Saha~Dalal}\ \emph {et~al.}(2012)\citenamefont
  {Saha~Dalal}, \citenamefont {Albaugh}, \citenamefont {Hoda},\ and\
  \citenamefont {Larson}}]{dala:12}%
  \BibitemOpen
  \bibfield  {author} {\bibinfo {author} {\bibfnamefont {I.}~\bibnamefont
  {Saha~Dalal}}, \bibinfo {author} {\bibfnamefont {A.}~\bibnamefont {Albaugh}},
  \bibinfo {author} {\bibfnamefont {N.}~\bibnamefont {Hoda}},\ and\ \bibinfo
  {author} {\bibfnamefont {R.~G.}\ \bibnamefont {Larson}},\ }\bibfield  {title}
  {\enquote {\bibinfo {title} {Tumbling and deformation of isolated polymer
  chains in shearing flow},}\ }\href@noop {} {\bibfield  {journal} {\bibinfo
  {journal} {Macromolecules}\ }\textbf {\bibinfo {volume} {45}},\ \bibinfo
  {pages} {9493} (\bibinfo {year} {2012})}\BibitemShut {NoStop}%
\bibitem [{\citenamefont {Martin-Gomez}, \citenamefont {Gompper},\ and\
  \citenamefont {Winkler}(2018)}]{mart:18}%
  \BibitemOpen
  \bibfield  {author} {\bibinfo {author} {\bibfnamefont {A.}~\bibnamefont
  {Martin-Gomez}}, \bibinfo {author} {\bibfnamefont {G.}~\bibnamefont
  {Gompper}},\ and\ \bibinfo {author} {\bibfnamefont {R.~G.}\ \bibnamefont
  {Winkler}},\ }\bibfield  {title} {\enquote {\bibinfo {title} {Active
  {B}rownian filamentous polymers under shear flow},}\ }\href@noop {}
  {\bibfield  {journal} {\bibinfo  {journal} {Polymers}\ }\textbf {\bibinfo
  {volume} {10}},\ \bibinfo {pages} {837} (\bibinfo {year} {2018})}\BibitemShut
  {NoStop}%
\bibitem [{\citenamefont {Winkler}(2006)}]{wink:06.1}%
  \BibitemOpen
  \bibfield  {author} {\bibinfo {author} {\bibfnamefont {R.~G.}\ \bibnamefont
  {Winkler}},\ }\bibfield  {title} {\enquote {\bibinfo {title} {Semiflexible
  polymers in shear flow},}\ }\href@noop {} {\bibfield  {journal} {\bibinfo
  {journal} {Phys. Rev. Lett.}\ }\textbf {\bibinfo {volume} {97}},\ \bibinfo
  {pages} {128301} (\bibinfo {year} {2006})}\BibitemShut {NoStop}%
\bibitem [{\citenamefont {Munk}\ \emph {et~al.}(2006)\citenamefont {Munk},
  \citenamefont {Hallatschek}, \citenamefont {Wiggins},\ and\ \citenamefont
  {Frey}}]{munk:06}%
  \BibitemOpen
  \bibfield  {author} {\bibinfo {author} {\bibfnamefont {T.}~\bibnamefont
  {Munk}}, \bibinfo {author} {\bibfnamefont {O.}~\bibnamefont {Hallatschek}},
  \bibinfo {author} {\bibfnamefont {C.~H.}\ \bibnamefont {Wiggins}},\ and\
  \bibinfo {author} {\bibfnamefont {E.}~\bibnamefont {Frey}},\ }\bibfield
  {title} {\enquote {\bibinfo {title} {Dynamics of semiflexible polymers in a
  flow field},}\ }\href@noop {} {\bibfield  {journal} {\bibinfo  {journal}
  {Phys. Rev. E}\ }\textbf {\bibinfo {volume} {74}},\ \bibinfo {pages} {041911}
  (\bibinfo {year} {2006})}\BibitemShut {NoStop}%
\bibitem [{\citenamefont {Kobayashi}\ and\ \citenamefont
  {Yamamoto}(2010)}]{koba:10}%
  \BibitemOpen
  \bibfield  {author} {\bibinfo {author} {\bibfnamefont {H.}~\bibnamefont
  {Kobayashi}}\ and\ \bibinfo {author} {\bibfnamefont {R.}~\bibnamefont
  {Yamamoto}},\ }\bibfield  {title} {\enquote {\bibinfo {title} {Tumbling
  motion of a single chain in shear flow: A crossover from {Brownian} to
  non-{Brownian} behavior},}\ }\href@noop {} {\bibfield  {journal} {\bibinfo
  {journal} {Phys. Rev. E}\ }\textbf {\bibinfo {volume} {81}},\ \bibinfo
  {pages} {041807} (\bibinfo {year} {2010})}\BibitemShut {NoStop}%
\bibitem [{\citenamefont {Bird}\ \emph {et~al.}(1987)\citenamefont {Bird},
  \citenamefont {Curtiss}, \citenamefont {Armstrong},\ and\ \citenamefont
  {Hassager}}]{bird:87}%
  \BibitemOpen
  \bibfield  {author} {\bibinfo {author} {\bibfnamefont {R.~B.}\ \bibnamefont
  {Bird}}, \bibinfo {author} {\bibfnamefont {C.~F.}\ \bibnamefont {Curtiss}},
  \bibinfo {author} {\bibfnamefont {R.~C.}\ \bibnamefont {Armstrong}},\ and\
  \bibinfo {author} {\bibfnamefont {O.}~\bibnamefont {Hassager}},\ }\href@noop
  {} {\emph {\bibinfo {title} {Dynamics of Polymer Liquids}}},\ Vol.~\bibinfo
  {volume} {2}\ (\bibinfo  {publisher} {John Wiley \& Sons},\ \bibinfo
  {address} {New York},\ \bibinfo {year} {1987})\BibitemShut {NoStop}%
\bibitem [{\citenamefont {Aust}, \citenamefont {Kr{\"o}ger},\ and\
  \citenamefont {Hess}(1999)}]{aust:99}%
  \BibitemOpen
  \bibfield  {author} {\bibinfo {author} {\bibfnamefont {C.}~\bibnamefont
  {Aust}}, \bibinfo {author} {\bibfnamefont {M.}~\bibnamefont {Kr{\"o}ger}},\
  and\ \bibinfo {author} {\bibfnamefont {S.}~\bibnamefont {Hess}},\ }\bibfield
  {title} {\enquote {\bibinfo {title} {Structure and dynamics of dilute polymer
  solutions under shear flow via nonequilibrium molecular dynamics},}\
  }\href@noop {} {\bibfield  {journal} {\bibinfo  {journal} {Macromolecules}\
  }\textbf {\bibinfo {volume} {32}},\ \bibinfo {pages} {5660} (\bibinfo {year}
  {1999})}\BibitemShut {NoStop}%
\bibitem [{\citenamefont {Hsieh}\ and\ \citenamefont {Larson}(2004)}]{hsie:04}%
  \BibitemOpen
  \bibfield  {author} {\bibinfo {author} {\bibfnamefont {C.-C.}\ \bibnamefont
  {Hsieh}}\ and\ \bibinfo {author} {\bibfnamefont {R.~G.}\ \bibnamefont
  {Larson}},\ }\bibfield  {title} {\enquote {\bibinfo {title} {Modelling
  hydrodynamic interaction in {Brownian} dynamics: {Simulation} of extensional
  and shear flows of dilute solutions of high molecular weight polystyrene},}\
  }\href@noop {} {\bibfield  {journal} {\bibinfo  {journal} {J. Rheol.}\
  }\textbf {\bibinfo {volume} {48}},\ \bibinfo {pages} {995} (\bibinfo {year}
  {2004})}\BibitemShut {NoStop}%
\bibitem [{\citenamefont {Winkler}(2016)}]{wink:16}%
  \BibitemOpen
  \bibfield  {author} {\bibinfo {author} {\bibfnamefont {R.~G.}\ \bibnamefont
  {Winkler}},\ }\bibfield  {title} {\enquote {\bibinfo {title} {Dynamics of
  flexible active {B}rownian dumbbells in the absence and the presence of shear
  flow},}\ }\href {https://doi.org/10.1039/C5SM02965A} {\bibfield  {journal}
  {\bibinfo  {journal} {Soft Matter}\ }\textbf {\bibinfo {volume} {12}},\
  \bibinfo {pages} {3737} (\bibinfo {year} {2016})}\BibitemShut {NoStop}%
\bibitem [{\citenamefont {Suma}\ \emph {et~al.}(2014)\citenamefont {Suma},
  \citenamefont {Gonnella}, \citenamefont {Laghezza}, \citenamefont {Lamura},
  \citenamefont {Mossa},\ and\ \citenamefont {Cugliandolo}}]{suma:14.1}%
  \BibitemOpen
  \bibfield  {author} {\bibinfo {author} {\bibfnamefont {A.}~\bibnamefont
  {Suma}}, \bibinfo {author} {\bibfnamefont {G.}~\bibnamefont {Gonnella}},
  \bibinfo {author} {\bibfnamefont {G.}~\bibnamefont {Laghezza}}, \bibinfo
  {author} {\bibfnamefont {A.}~\bibnamefont {Lamura}}, \bibinfo {author}
  {\bibfnamefont {A.}~\bibnamefont {Mossa}},\ and\ \bibinfo {author}
  {\bibfnamefont {L.~F.}\ \bibnamefont {Cugliandolo}},\ }\bibfield  {title}
  {\enquote {\bibinfo {title} {Dynamics of a homogeneous active dumbbell
  system},}\ }\href {https://doi.org/10.1103/PhysRevE.90.052130} {\bibfield
  {journal} {\bibinfo  {journal} {Phys. Rev. E}\ }\textbf {\bibinfo {volume}
  {90}},\ \bibinfo {pages} {052130} (\bibinfo {year} {2014})}\BibitemShut
  {NoStop}%
\bibitem [{\citenamefont {Kapral}(2008)}]{kapr:08}%
  \BibitemOpen
  \bibfield  {author} {\bibinfo {author} {\bibfnamefont {R.}~\bibnamefont
  {Kapral}},\ }\bibfield  {title} {\enquote {\bibinfo {title} {Multiparticle
  collision dynamics: Simulations of complex systems on mesoscale},}\
  }\href@noop {} {\bibfield  {journal} {\bibinfo  {journal} {Adv. Chem. Phys.}\
  }\textbf {\bibinfo {volume} {140}},\ \bibinfo {pages} {89} (\bibinfo {year}
  {2008})}\BibitemShut {NoStop}%
\end{thebibliography}
\end{document}